\IfFileExists{arxiv.tex}{\input{arxiv.tex}} 
\newcommand{\arxiv}{}

\documentclass{sig-alternate}

\usepackage{mathptmx} %
\usepackage{fancyhdr}
\usepackage{multirow}
\usepackage[table,dvipsnames,usenames]{xcolor}
\usepackage{amsmath,amssymb}
\usepackage{xspace}
\usepackage{subfig}
\usepackage[margin=0cm]{caption}
\usepackage{textcomp}
\usepackage{svg}
\usepackage{calc}
\usepackage{tabularx}
\usepackage{soul}
\usepackage{marginnote}
\usepackage{mdframed}
\usepackage{textcomp}
\usepackage{enumitem}
\usepackage{url}
\usepackage{svg}
\usepackage{enumitem}

\usepackage[bookmarks=true,breaklinks=true,letterpaper=true,colorlinks,citecolor=blue,linkcolor=blue,urlcolor=blue]{hyperref}
\usepackage[normalem]{ulem}

\ifdefined\arxiv
  \author{Suyash Mahar$^{*}$, Hao Wang$^{\dagger}$, Wei Shu$^{\mathsection}$, Abhishek Dhanotia$^{\ddagger}$\\
  $^{*}$UC San Diego\quad{}$^{\dagger}$NVIDIA\quad{}$^{\mathsection}$Tenstorrent\quad{}$^{\ddagger}$Meta Inc.}
\else
  \newcommand{\microsubmissionnumber}{568}
  \fancypagestyle{firstpage}{
    \fancyhf{}
    
    \fancyhead[C]{\vspace{10pt}\normalsize{MICRO 2023 Submission
        \textbf{\#\microsubmissionnumber} -- Confidential Draft -- Do NOT Distribute!!}\\\vspace{-25pt}} 
    \fancyfoot[C]{\thepage}
  }
\fi

\newcommand{\chpar}[1]{#1}

\newcommand{\AB}[1]{\abhishek{#1}}

\newcommand{\ignore}[1]{}

\newcommand{\reffig}[1]{Figure~\ref{#1}}%
\newcommand{\refsec}[1]{Section~\ref{#1}}%
\newcommand{\reftab}[1]{Table~\ref{#1}}%
\newcommand{\reflns}[2]{Lines~\hyperref[#1]{\ref*{#1}-\ref*{#2}}}%

\newcommand{\x}[1]{$\times$}

\newcommand{\web}{Web1\xspace}
\newcommand{\insta}{Web2\xspace}
\newcommand{\adfinder}{Ads1\xspace}
\newcommand{\adranker}{Ads2\xspace}
\newcommand{\adretriever}{Ads3\xspace}
\newcommand{\memcache}{Cache1\xspace}
\newcommand{\tao}{Cache2\xspace}
\newcommand{\dppreader}{Reader\xspace}
\newcommand{\feed}{Feed\xspace}

\newcommand{\itlb}{I-TLB\xspace}
\newcommand{\dtlb}{D-TLB\xspace}
\newcommand{\memprof}{MemProf\xspace}
\newcommand{\memprofcode}{MemProf.Code\xspace}
\newcommand{\memprofbw}{MemProf.MemBW\xspace}
\newcommand{\memproflat}{MemProf.MemLat\xspace}

\newcommand{\doublerown}[1]{\multirow{-2}{*}{\shortstack{#1}}}
\newcommand{\triplerown}[1]{\multirow{-3}{*}{\shortstack{#1}}}

\newcommand{\smahar}[1]{{{\textcolor{Blue}{\textbf{\sffamily
          $\blacktriangleright$ Suyash: #1 $\blacktriangleleft$}}}}}
\newcommand{\abhishek}[1]{{{\textcolor{red}{\textbf{\sffamily
          $\blacktriangleright$ Abhishek: #1 $\blacktriangleleft$}}}}}
\newcommand{\wei}[1]{{{\textcolor{purple}{\textbf{\sffamily
          $\blacktriangleright$ Wei: #1 $\blacktriangleleft$}}}}}
\newcommand{\hao}[1]{{{\textcolor{orange}{\textbf{\sffamily
          $\blacktriangleright$ Hao: #1 $\blacktriangleleft$}}}}}

\newcommand{\tracing}[1]{{#1}}

\mdfdefinestyle{observation}{leftmargin=0cm,rightmargin=0cm,%
innerleftmargin=0.2cm,innerrightmargin=0.2cm,roundcorner=10pt}
\newcommand{\obs}[2]{\refstepcounter{obscounter}%
    \noindent{}\begin{mdframed}[backgroundcolor=gray!10,style=observation]
        \textbf{Key observation \theobscounter\label{#1}:} #2
    \end{mdframed}
}

\newcommand{\uarch}{\textmu{}-arch\xspace}
\soulregister\uarch7

\ifdefined\arxiv
  \newcommand{\company}{Meta\xspace}
\else
  \newcommand{\company}{a hyperscalar\xspace}
\fi

\def\shipit{1}

\soulregister\itlb7

\soulregister\AB7
\soulregister\abhishek7
\soulregister\web7
\soulregister\insta7
\soulregister\adfinder7
\soulregister\adranker7
\soulregister\adretriever7
\soulregister\memcache7
\soulregister\tao7
\soulregister\dppreader7
\soulregister\feed7
\soulregister\reffig7
\soulregister\refsec7
\soulregister\reftab7

\if\shipit1
    \renewcommand{\hl}[1]{#1}
    \renewcommand{\marginpar}[1]{}
    \renewcommand{\smahar}[1]{}
    \renewcommand{\abhishek}[1]{}
    \renewcommand{\wei}[1]{}
    \renewcommand{\hao}[1]{}
\fi

\title{Workload Behavior Driven Memory Subsystem Design for Hyperscale}

\begin{document}

\newcounter{obscounter}

\date{}
\maketitle

\ifdefined\arxiv
\else
  \thispagestyle{firstpage}
\fi

\begin{abstract}
\vspace{-0.5cm}
Hyperscalars run services across a large fleet of servers, serving billions of users worldwide.
However, these services exhibit different behaviors compared to commonly available benchmark suites, leading to server architectures that are suboptimal for cloud workloads.
As datacenters emerge as the primary server processor market, optimizing server processors for cloud workloads by better understanding their behavior is an area of interest.
To address this, we present \memprof, a memory profiler that profiles the three major reasons for stalls in cloud workloads: code-fetch, memory bandwidth, and memory latency. We use \memprof to understand the behavior of cloud workloads at \company and propose and evaluate micro-architectural and memory system design improvements that help cloud workloads' performance.

\memprof's code analysis shows that cloud workloads at \company execute the same code across CPU cores. Using this, we propose shared micro-architectural structures--a shared L2 \itlb and a shared L2 cache\ignore{ to make the apparent code cache size larger without significantly increasing its area cost}. Next, to help with memory bandwidth stalls, using \ignore{ we measure} workloads' memory bandwidth distribution, we find that only a few pages contribute to most of the system bandwidth. We use this finding to evaluate a new high-bandwidth, small-capacity memory tier and show that it performs  1.46$\times$ better than the current baseline configuration.
Finally, we look into ways to improve  memory latency for cloud workloads. Profiling using \memprof reveals that L2 hardware prefetchers, which are commonly used to reduce memory latency, \ignore{are meant to help hide memory latency, they} have very low coverage and consume a significant amount of memory bandwidth. To help improve future hardware prefetcher performance, we built an efficient memory tracing tool to collect and validate production memory access traces. 
Our memory tracing tool adds significantly less overhead than DynamoRIO, enabling tracing production workloads.
\end{abstract}
\vspace{-0.1cm}

\section{Introduction}
\label{sec:intro}

In recent years, the market share of the cloud as a portion of the total server market has constantly been increasing. Cloud deployments now account for >50\% of the total server processor market and are expected to grow even further~\cite{datacenter0, datacenter1}. However, CPU benchmarks available today \ignore{to optimize these processors}do not accurately represent microarchitectural and memory system behavior of real cloud workloads, leaving the processors largely unoptimized for their unique characteristics.

Cloud workloads show different behavior than the CPU benchmarks available today across several metrics. For one, the fleet-wide\footnote{Server fleet refers to an ensemble of servers across datacenters.} IPC (instructions per cycle) for major hyperscalers is significantly lower than that of the benchmark suites~\cite{sriraman2019softsku,kanev2015profiling}. With millions of servers running these workloads, even small improvements in IPC from understanding the behavior of these workloads can result in significant cost savings and efficiency improvements across the fleet.

To improve the performance of cloud workloads, in this paper, we study and characterize nine microservices serving live production traffic. These microservices represent a diverse range of workloads at a hyperscalar’s datacenters and run on a significant portion of the fleet.

\begin{figure}
    \centering
    \includegraphics[width=0.95\linewidth]{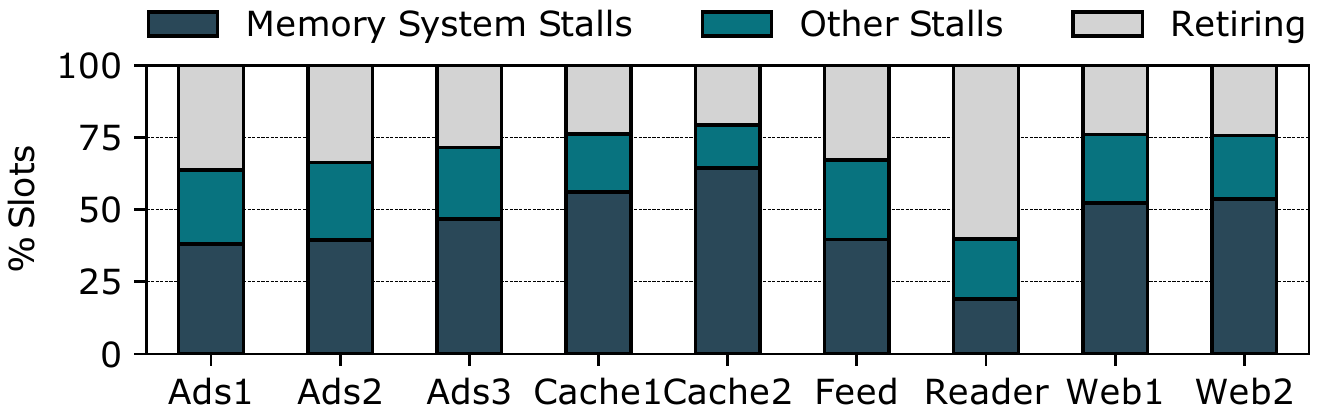}
    \vspace{-0.15cm}
     \caption{Breakdown of where the CPU pipeline slots spend their time across nine production workloads using top-down analysis~\cite{yasin2014top}.}
    \label{fig:mem-bound}
\end{figure}

We find that the memory system is the primary reason for CPU pipeline stalls across most of these cloud workloads, as shown in \reffig{fig:mem-bound} using the top-down analysis~\cite{yasin2014top}.
These memory system-related stalls can have three sources, (1) stall on an instruction fetch (\textit{code-fetch bound}), (2) stall on the memory system with a high request pending queue occupancy (\textit{memory bandwidth bound}), or (3) stall on the memory system with a low request pending queue occupancy (\textit{memory latency bound}).

To further dive into these three sources of stalls in cloud workloads, we created a new tool, \memprof, that enables detailed profiling of workloads' micro-architectural and memory system behavior. We use observations from \memprof to propose and evaluate new micro-architectural changes and system architecture for cloud server processors.

The first source of stalls, code fetch-related stalls are a significant problem in cloud workloads~\cite{ayers2018memory, sriraman2019softsku}. Using \memprof, we observe that workloads at \company run similar code and share page table mappings across cores. Building on this observation, we propose sharing the L2 Cache and L2 \itlb among a small cluster of  cores (e.g., four) that are physically close. Shared \textmu-arch structures for instructions help reduce code-fetch stalls by enabling the cores running similar code to pool their caches for larger apparent cache size, even though they have the same per-core cache size.

Memory bandwidth is another significant contributor to memory system stalls. Over the years, DDR memory capacity and bandwidth have been scaling at significantly different rates ~\cite{maruf2022tpp}. This increasing memory capacity and bandwidth disparity, along with growing core counts in modern server processors has resulted in limited per-core memory bandwidth, requiring processors to populate more memory channels, thus resulting in higher server TCO (total cost of ownership).  We use \memprof to study the memory bandwidth distributions of production workloads and find that only a small percentage of pages contribute to most of the memory bandwidth utilization across the workloads. We use this observation to propose and evaluate new \textit{bandwidth-tiered} memory systems as opposed to latency-tiered systems common today. In this new system, a small, high-bandwidth memory tier serves the bandwidth needs of the workload. In contrast, a low-bandwidth, large-capacity tier provides additional memory capacity, lowering servers' TCO. We find that by splitting the memory capacity in 30:70\% between the high-BW and low-BW tiers, we can achieve 1.46$\times$ better throughput and 13\% better throughput/cost than the baseline DDR-only configuration.

Finally, we look into ways to improve future memory latency research for cloud workloads. 
Using \memprof, we observe that although L2 hardware prefetchers improve memory latency by prefetching cachelines, they often have very low coverage in real-world cloud workloads, resulting in minor IPC improvements. 
Further, we observe that L2 hardware prefetcher's inefficiencies result in high memory bandwidth consumption, exacerbating the memory bandwidth problem.
A major hurdle in optimizing these hardware prefetchers is access to the cloud workloads to study their memory behavior. To enable broader access to cloud workloads' behavior, we \tracing{built an efficient memory tracing tool capable of tracing live production workloads and} collect several memory access traces of live production workloads. \tracing{Our tracing tool has significantly less runtime overhead compared to state-of-the-art, DynamoRIO.} \tracing{We} plan to make the traces available in the future to help  hardware prefetcher research.

This paper is the first to study how cloud workloads interact with the memory system and to use these observations to propose shared \uarch{} structures and memory bandwidth tiering. While previous works from Google~\cite{ayers2018memory, kanev2015profiling} and Meta~\cite{sriraman2019softsku, sriraman2020accelerometer} have investigated the \uarch{} behavior of hyperscale workloads, they have not performed detailed studies of the code behavior, memory bandwidth distribution, or hardware prefetcher efficiency. In summary, this paper makes the following contributions:
\begin{itemize}[leftmargin=15pt, rightmargin=0cm,itemsep=-1pt]
    \item \textbf{Detailed study and characterization of cloud workloads.} We perform a detailed study of code, memory bandwidth, and memory latency trends at \company. 
    \item \textbf{New profiling methodology and tool.} We present \memprof, a profiling tool for cloud workloads that helps understand their interaction with the memory system.
    \item \textbf{Code sharing across cores.} We show that CPU cores run very similar code across all the workloads, enabling performance improvements from shared \uarch{} resources.
    \item \textbf{Memory bandwidth distribution.} We study the memory bandwidth distribution, find that very few pages contribute to a large amount of bandwidth and propose new tiered memory solutions for cloud workloads.
    \item \textbf{Hardware prefetcher efficiency.} L2 hardware prefetchers incur significant memory bandwidth overhead with small performance improvements. To address this, we will make production workload memory traces available to the community for future hardware prefetcher research.
\end{itemize}

\noindent{}In the next section, we look into the three reasons for memory system stalls: code fetch, memory bandwidth, and memory latency. We follow with the description of \memprof{} and use it to make observations and proposals in each area.

\ignore{\noindent{}In the next section, we look into the memory system stalls to understand the contribution of code fetch, memory bandwidth, and memory latency. Next, we present \memprof in \refsec{sec:memprof}, followed by the observations and proposals for improving code fetch-related stalls (\refsec{sec:code-fetch}), memory bandwidth-related stalls (\refsec{sec:mem-bw}), and memory latency-related stalls (\refsec{sec:mem-lat}). Finally, we discuss related works in \refsec{sec:related} and the conclusion in \refsec{sec:conclusion}. }
\section{Code, Memory BW, and Latency Challenges in Cloud Datacenters}
\label{sec:motivation}

Existing benchmarks that are generally used to design processors do not accurately represent the workloads at hyperscalars. As highlighted by several previous works~\cite{ayers2018memory,ayers2019asmdb,kanev2015profiling, sriraman2019softsku, sriraman2020accelerometer},  cloud workloads exhibit fundamentally different IPC, cache miss rates, and other metrics compared to standard benchmarks. These factors make the SPEC and other commercially available benchmarks a bad proxy for studying the performance of server processors in a datacenter.

\begin{figure}
    \centering
        \includegraphics[width=\linewidth]{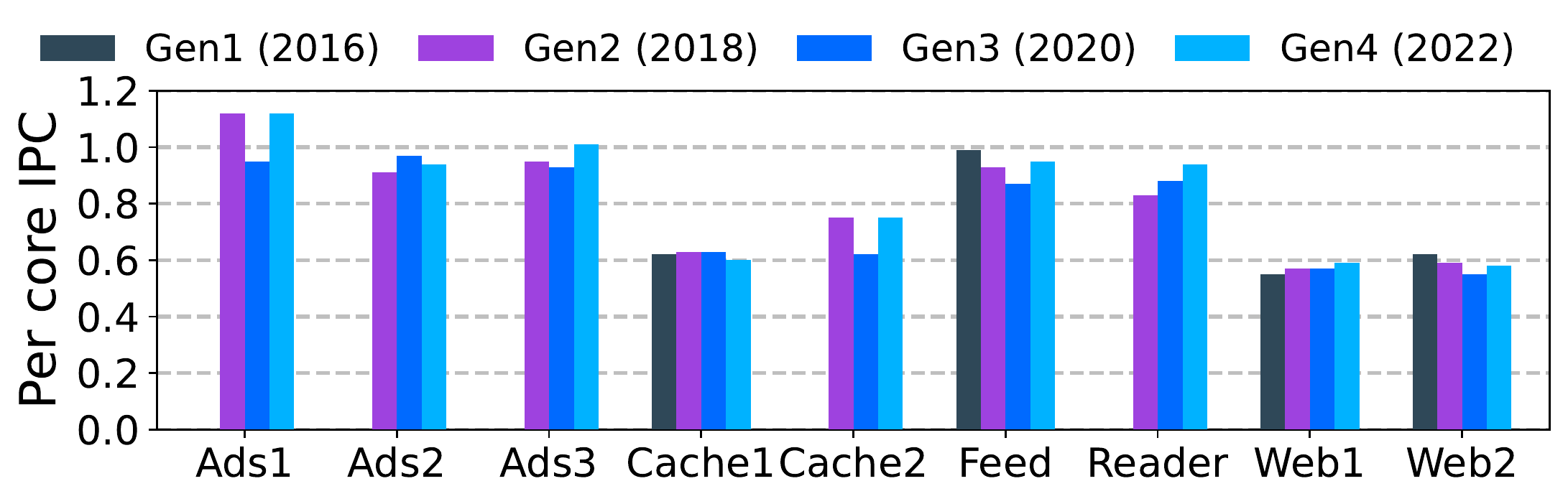}
    \vspace{-0.6cm}
    \caption{Gen-over-gen IPC change for the 9 workloads. \tracing{Missing Gen1 data corresponds to services that do not run on Gen1.} Years correspond to processor release year. %
    }
    \label{fig:gen-over-gen-ipc}
\end{figure}

Besides standard benchmarks' little resemblance to cloud workloads, we motivate the need to better understand cloud workloads by comparing the IPC change across server generations at \company. We chart the IPC change over the years across nine representative cloud workloads and four server generations running the same software versions. \reffig{fig:gen-over-gen-ipc} shows that the IPC improvement is minimal or even negative across generations, necessitating the need for a deeper understanding of cloud workloads.

To understand the reasons for this inhibited performance growth of server processors running cloud workloads, we will look into each of the three primary reasons for memory system stalls \ignore{from \reffig{fig:emon-trend}}and understand how they affect cloud workloads' performance.

\subsection{Increasing Code Footprint of Datacenter Applications}
\begin{figure}
    \centering
    \includegraphics[width=\linewidth]{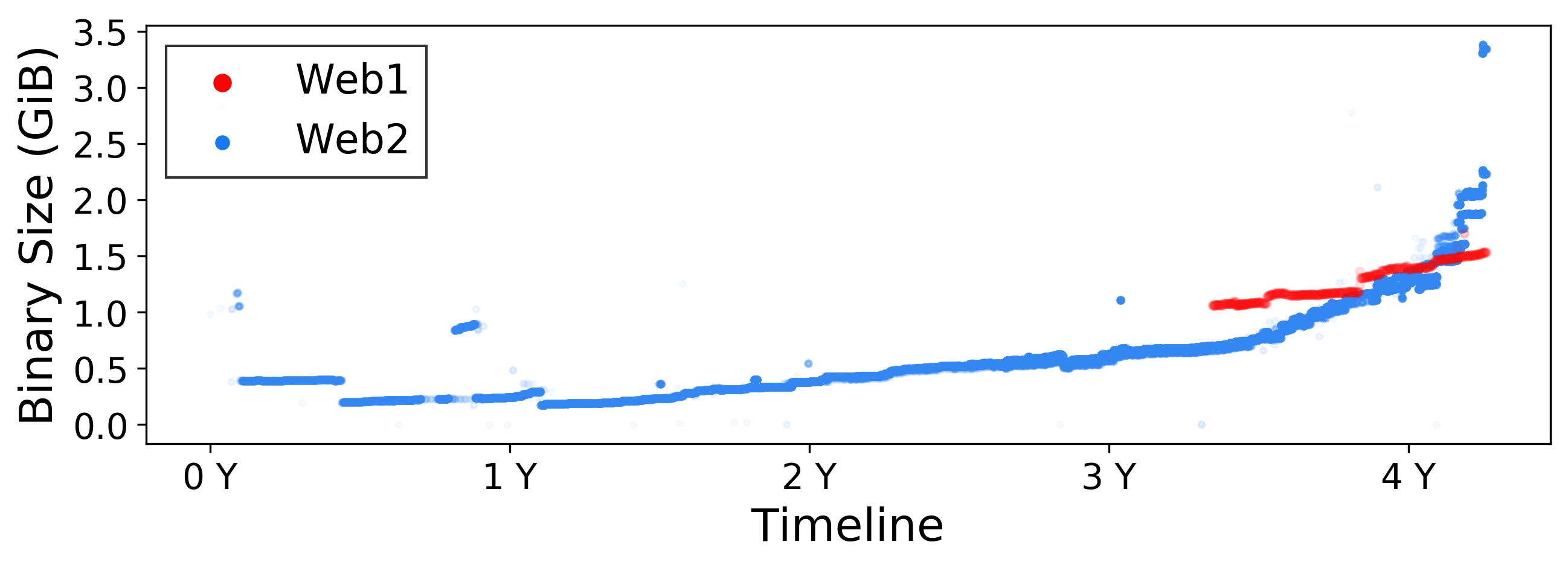}
    \vspace{-0.5cm}
    \caption{Binary Size trend for \web and \insta. \insta binary size increased by 5.6$\times$ over the last four years.}
    \label{fig:codesize-trend}
\end{figure}

Cloud workloads are often frontend bound due to their large code footprint, resulting in high instruction cache miss rates~\cite{ayers2018memory}. The code footprint can be hundreds of MiBs and has been increasing over the years. \reffig{fig:codesize-trend} shows the increase in the binary size for \web and \insta, the two major web services at \company over the years. \insta shows an exponential increase in binary size, a 5.6$\times$ increase over four years. 

In contrast, the I-cache sizes have stayed relatively constant (32KB I\$ per core) across several server generations and have only recently started increasing on some server architectures. \marginpar{for the generations, they haven't; Hao modified}
Likewise, instruction TLB lookups are optimized for the latency-sensitive instruction fetch pipeline, making it harder for them to scale with cloud workloads' rapidly growing code footprints.
Previous work on cloud workload has observed a similar code-footprint trend. Kanev et al.~\cite{kanev2015profiling} show that the code footprint at Google is growing at a rate of 27\% per year. This imbalance between the processor's micro-architectural resources and the growing code footprint have resulted in frontend to be the leading source of bottleneck for many workloads across several server generations and resulted in stagnant IPC over time.

\subsection{Memory Bandwidth Scaling Challenge}\label{sec:mem-bw-motivation}
\begin{figure}
    \centering
    \includegraphics[width=\linewidth]{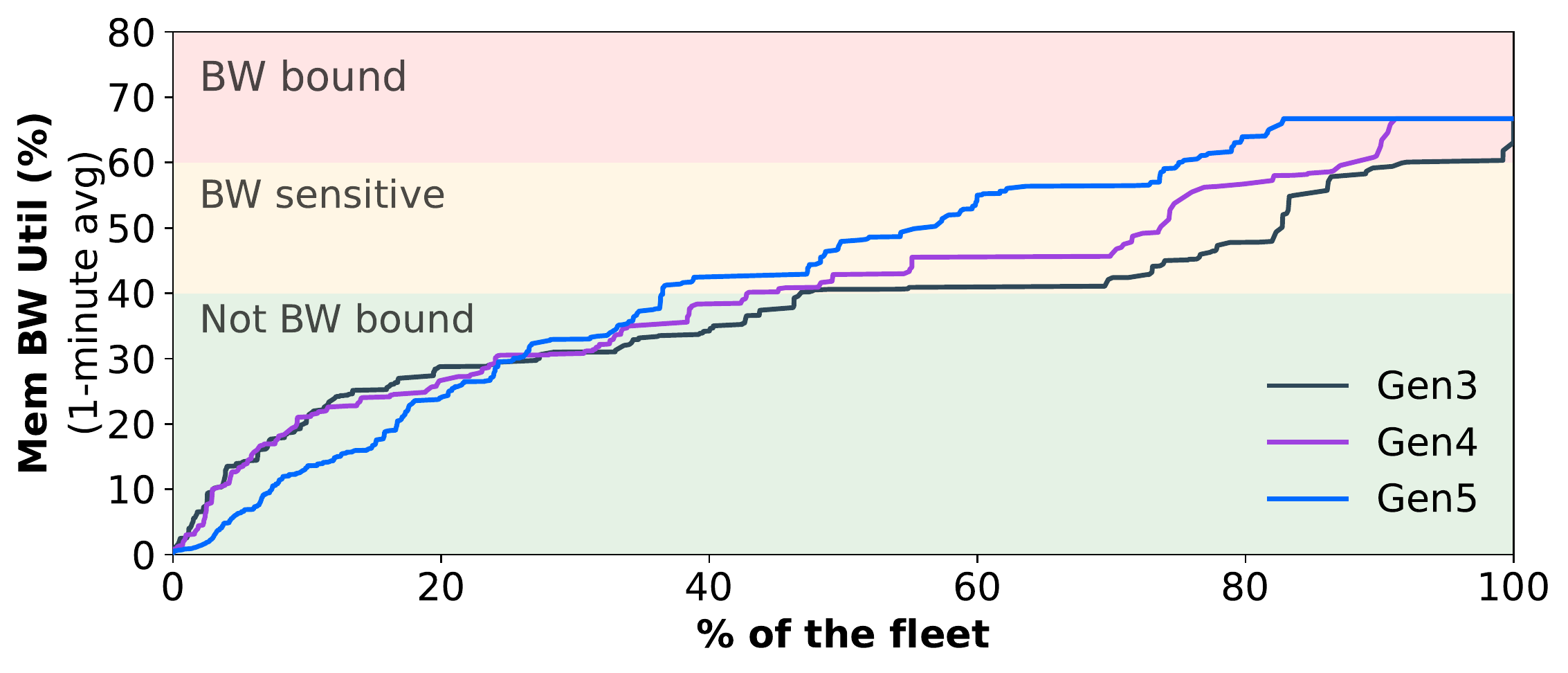}
    \caption{
    Memory bandwidth utilization (\%) CDF across three generations (1-min avg); BW bound and sensitive regions based on DDR memory bandwidth vs. latency characteristics.
    }
    \label{fig:fleet-mem-bw}
\end{figure}

Memory bandwidth is another major concern for cloud workloads, resulting from poor DRAM bandwidth scaling and increasing CPU core count. To understand cloud workloads' memory bandwidth utilization trend, we show the fleet-wide memory bandwidth consumption increase over the three generations of servers in \reffig{fig:fleet-mem-bw}. With the most recent generation (Gen 5), the 1-minute average memory bandwidth utilization shows that the majority of the fleet is either bandwidth sensitive or bandwidth bound. 

While mirco-benchmarks can generally drive memory bandwidth utilization to >80\%, we observe that production workloads rarely exceed 60\% memory bandwidth utilization, as any further increase results in an exponential increase in memory latency~\cite{radulovic2015another}. Thus, we classify workloads with higher than 60\% bandwidth utilization as memory bandwidth bound (shaded red in \reffig{fig:fleet-mem-bw}). Likewise, workloads with average memory bandwidth utilization between 40\%-60\% can have high transient memory bandwidth utilization and are thus classified as memory BW sensitive. While the fleet-wide data is 1-minute average, transient peaks can cause these workloads to become significantly bandwidth bound during shorter time intervals, causing tail latency spikes and limited overall system CPU utilization.

Besides growing memory bandwidth utilization across the fleet, increasing disparity between DIMM's bandwidth and capacity~\cite{maruf2022tpp} makes alleviating memory bandwidth problems of cloud workloads harder as hyperscalars need to provision a large number of memory channels and DIMMs per server for sufficient memory bandwidth. 

The trend of increasing memory bandwidth utilization and the poor bandwidth scaling of DDR-based memories call for more efficient memory provisioning strategies in cloud datacenters to reduce server costs and power consumption~\cite{dayarathna2015data}.

\subsection{High Memory Latency Leading to CPU Underutilization}

\begin{figure}
    \centering
    \includegraphics[width=\linewidth]{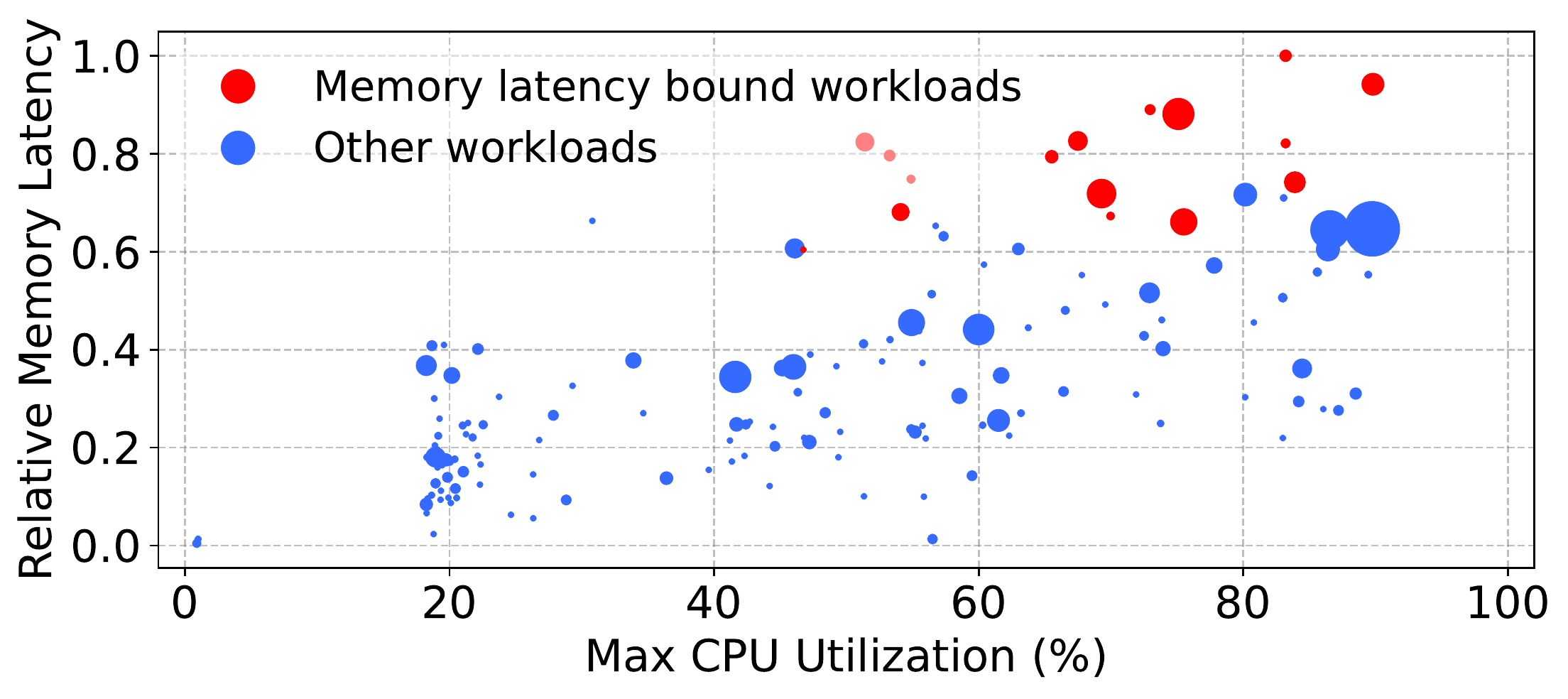}
    \caption{
    Relative memory latency vs. max CPU utilization for microservices; red markers indicate latency SLO constraints limiting CPU utilization; marker size represents service size; latency relative to peak load.}
    \label{fig:latency-vs-cpu-utilization}
        \vspace{-0.15cm}
\end{figure}

Finally, many cloud workloads are memory latency sensitive as they must meet service level objectives (SLOs) for request latencies~\cite{ding2019characterizing}. These workloads run at the maximum CPU utilization possible without violating SLO guarantees, often wasting CPU resources. \hl{Memory latency limits CPU utilization as an increase in memory latency lowers the system's IPC. Lower IPC for workloads results in cloud workloads spending more cycles doing ``on-cpu'' work, and thus results in longer query response times.}

To study this trend across the fleet, \reffig{fig:latency-vs-cpu-utilization} shows \tracing{a scatter plot with} the max CPU utilization vs. memory latency of workload relative to latency at peak load. Peak load is the max throughput a workload can
sustain without violating any SLOs. \hl{We make three important observations from the fleet-wide data: (a) Memory latency increases significantly with increasing CPU utilization. That is, a large portion of the fleet is operating at a relatively high memory latency (>0.5). (b) Workloads marked in red have 20-50\% stranded CPU cores because of the latency SLO constraints.} These workloads must leave CPU cores idle to avoid violating SLO guarantees\ignore{ resulting from high memory latency}, resulting in wasted resources and inefficiency at the datacenter scale. (c) And differences in memory access patterns (e.g., bursty memory load) result in workloads experiencing varying memory latencies at similar CPU utilization levels. %

Hardware prefetching is one of the key techniques used in modern server processors to hide memory latency. L2 hardware prefetchers, however, have a significant memory bandwidth overhead. 
Our analysis in \refsec{sec:mem-lat} reveals that workloads show a substantial increase in total memory bandwidth consumption with L2 prefetchers enabled while providing limited performance gains, signaling that the prefetchers have low efficiency.

Next, we'll introduce \memprof, a profiling tool to characterize cloud workloads and explore performance improvement opportunities in cloud server processors.

\section{\memprof}\label{sec:memprof}
So far, our focus has been on the sources and the consequences of memory system stalls. In order to address these issues, we present \memprof, a code and memory profiler, to profile and understand the memory subsystem behavior of cloud workloads.  \memprof is an automated tool that measures several hardware counters, samples events using Intel PEBS~\cite{intel-pebs}, and generates reports on code sharing, memory bandwidth distribution, and memory latency behavior.

\subsection{Design}

\begin{figure*}
    \centering
    \includegraphics[width=\linewidth]{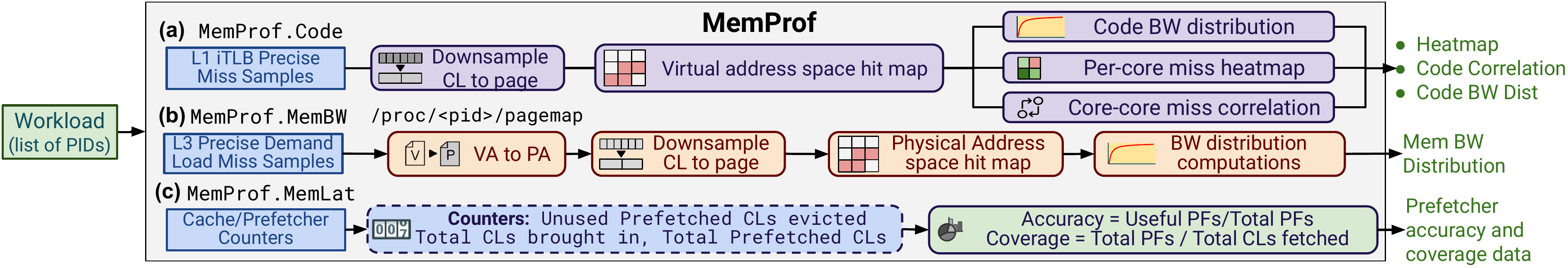}
    \caption{\memprof overview. (a) \memprofcode, (b) \memprofbw, (c) \memproflat profile the code, memory BW, and memory latency behavior, respectively.}
    \label{fig:memprof-overview}
    \vspace{-0.6cm}
\end{figure*}

\reffig{fig:memprof-overview} shows the overview of \memprof. \memprof has three components that measure the three aspects of memory system-related stalls. (1) \textit{\memprofcode} measures a workload's code behavior  to understand code sharing across cores and code bandwidth distribution. (2) \textit{\memprofbw} generates the memory bandwidth distribution to help understand how hot memory pages in a system are. And, (3) \textit{\memproflat} generates data to understand the effectiveness of common techniques in reducing memory latency. %

\subsection{Methodology} 

Next, we will look at the methodology used by \memprof and understand its working.

\vspace{4pt}\noindent\textbf{\memprofcode:} For profiling the code behavior of workloads, \memprof{} samples per-core L1 I-TLB miss events (\texttt{FRONTEND\_RETIRED.ITLB\_MISS}) using Intel PEBS. Larger granularity sampling using I-TLB miss events enables \memprof to cover wider address space regions. Further, I-TLB-based miss sampling  and dense binary layout of hot code using link-time optimization~\cite{bolt-anonymous} enable low frequency, low-overhead code footprint estimation. 

With the I-TLB miss event samples, \memprof computes the hit distribution across the address space of the workload and uses it to generate the code heatmap, code bandwidth distribution, and compute the correlation between the miss events of two cores (\reffig{fig:memprof-overview}a). 
 For all these calculations, \memprof uses the virtual address of the samples.

\memprofcode's results help understand the code footprint and code sharing across cores in cloud workloads.

\vspace{4pt}\noindent\textbf{\memprofbw:} To understand the memory bandwidth requirements of a workload, \memprof samples the precise LLC demand load miss event (\texttt{MEM\_LOAD\_RETIRED.L3\_MISS}) and generates the memory bandwidth distribution profile for a workload (\reffig{fig:memprof-overview}b). 
As multiple virtual pages can map to the same physical page, \memprof translates sampled virtual addresses to corresponding physical addresses using the workload's page table before computing the memory bandwidth distribution.
\memprof supports sampling over a configurable interval to study memory bandwidth trends over time and translates the samples' address at the end of the sampling interval. In cases where \ignore{the application creates or deletes mappings frequently, }the page table entries for an address range might not exist at the end of a measurement interval\ignore{. To solve this}, \memprof keeps around the last copy of the page table and uses it to translate any addresses missing in the current copy.

\vspace{4pt}\noindent\textbf{\memproflat:} 
To measure how effective hardware prefetchers are in reducing memory latency, \memprof performs two sets of measurements. To compute their accuracy and coverage, \memprof measures the cache and prefetcher-related hardware counters and uses the following expressions. (CL = cachelines, pref. = prefetched)
\vspace{-0.2cm}\newcommand{\mathtok}[1]{\textit{#1}}
\begin{align*}
    &\mathtok{Accuracy} = 1-\frac{(\mathtok{Unused pref. CLs evicted})}{(\mathtok{Total pref. CLs})}\\
    &\mathtok{Coverage} ={}\\
    & \frac{(\mathtok{Total pref. CLs}) - (\mathtok{Unused pref. CLs evicted})}{(\mathtok{Total CLs brought in}) - (\mathtok{Unused pref. CLs evicted})}
\end{align*}

To measure the memory bandwidth overhead of enabling L2 hardware prefetchers, \memprof measures the total system bandwidth consumption and IPC with L2 prefetchers enabled and disabled across all cores. Since these counters are only available for L2 caches, \memprof only generates accuracy and coverage data for L2 prefetchers. 

Collecting accuracy, coverage, and memory bandwidth overhead of L2 hardware prefetchers gives us insights into prefetcher behavior of different workloads.

\subsection{Production Servers and Workloads}

\begin{figure}
    \centering
    \includegraphics[width=0.9\linewidth]{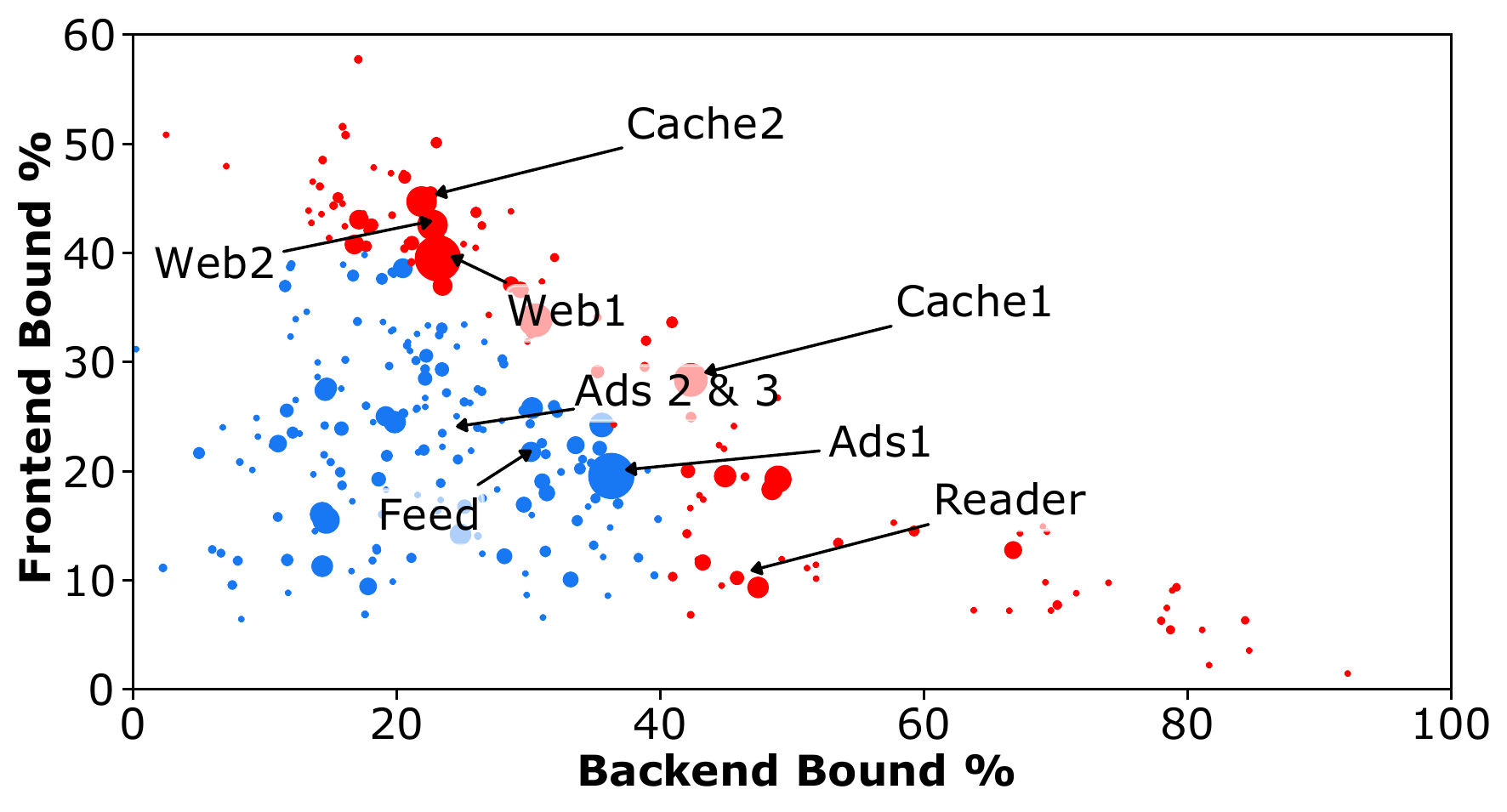}
    \vspace{-0.25cm}
    \caption{Distribution of workloads across the fleet based on \% frontend and backend stalls. Marker size is scaled by the workload's server count.}
    \label{fig:fb-vs-bb}
\end{figure}

\begin{table}
\fontsize{8}{10}
\selectfont
  \caption{Gen 4 system configuration.}
  \vspace{-0.2cm}
  \label{tab:gen4-config}
  \centering
  \rowcolors{2}{gray!25}{white}
  \begin{tabular}{|l|l|}
    \rowcolor{gray!25}
    \hline
    \rowcolor[HTML]{d8d8ff}
    \textbf{Parameter} & \textbf{Value}\\
    \hline
    Cores/Threads & 26/52 \\\hline
    L1 I-/D-Cache               & 32~KiB/core each, private \\\hline
    L2 Cache                     & 1~MiB/core, unified, private                  \\\hline
                &  2~MiB/4~MiB pages fully assoc., 8 entries/thread;  \\
                \rowcolor{gray!25}
    \doublerown{I-L1 TLB} & 4~KiB pages 8-way, 128 entries/core\\\hline
    \rowcolor{white}
    LLC                          & 1.375~MiB/core                       \\\hline
  \end{tabular}
\end{table}

We run containerized production workloads on bare metal machines
(that is, no hypervisor or virtual machines) to characterize
microservices in~\refsec{sec:memprof}, and for studies in~\refsec{sec:code-fetch} and~\ref{sec:mem-bw}. These workloads process live production traffic, represent a
large part of the fleet, and exhibit a wide range of
\textmu-architectural behaviors. We run each of these workloads near the peak load
operating point.

Further, all experiments were run on the fourth generation of servers, with the
hardware configuration listed in \reftab{tab:gen4-config}.

We use nine representative workloads running live production traffic to
study the frontend and backend behavior of the fleet.  \reffig{fig:fb-vs-bb}
shows the percentage of CPU pipeline slots across the microservices that are
frontend and backend bound. \hl{Across the fleet, we see that workloads show
  extremely diverse \uarch{} behavior, ranging from highly frontend bound to
  highly backend bound.}

The nine labeled microservices in the scatter plot account for a significant portion of the fleet  and exhibit diverse \uarch{} behavior. We study these nine workloads to perform a representative characterization of cloud workloads at \company.%

\ignore{\paragraph{\web and \insta.} \web and \insta are the two web-serving workloads. \web uses a hip-hop virtual machine (HHVM)~\cite{adams2014hiphop} to serve web requests. While \insta is a Django-based web server that uses an optimized version of CPython~\cite{cpython} to interpret python bytecode. \insta{}'s optimized CPython implementation has its garbage collector disabled to improve requests' tail latency.

\paragraph{Ads1, Ads2, and Ads3.} \adfinder{}, \adranker{}, and \adretriever{} are the three ad microservices that maintain user-specific and ad-specific data. Together these services find, index, and retrieve ads to show them to the user.

\paragraph{\memcache and \tao.} 
\memcache and \tao are two in-memory caching microservices. \memcache implements a look-aside, while \tao implements a look-through cache. \memcache is bottlenecked by packet processing. To achieve the maximum query rate for \memcache, hosts use 2/3 of the cores for running the workload. The remaining 1/3 cores are used for network processing and handling NIC IRQs.

\paragraph{Feed}
Feed workload aggregates responses from other microservices into ``stories'', and then characterizes them using various models and feature extractors. These stories are then sent for scoring and ranking to display to the user.

\paragraph{Reader} The reader microservice is part of the AI training pipeline, where it reads the data from the storage cluster and preprocesses it for the AI training cluster~\cite{dppreader-anonymous}.}

These microservices are \web, \insta, \adfinder, \adranker, \adretriever, \memcache, \tao, \feed{}, and \dppreader{}. To achieve the maximum query rate for \memcache, hosts use 2/3 of the cores for running the workload. The rest of the 1/3 cores are used for network processing and handling NIC IRQs. Most microservices spawn multiple threads, where only one thread is assigned to a logical core. In contrast, \insta spawns multiple processes and assigns up to one process per logical core.

\section{Code Behavior of Cloud Workloads}\label{sec:code-fetch}

\begin{figure}
    \centering
    \includegraphics[width=\linewidth]{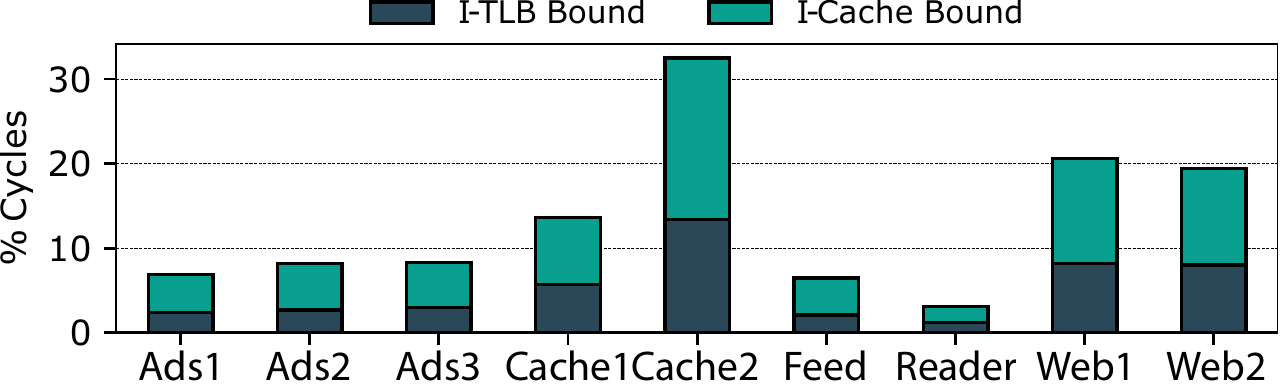}
    \caption{\% code fetch bound cycles (I-cache and I-TLB).}
    \label{fig:i-bound}
\end{figure}

\begin{figure*}
    \centering
    \includegraphics[width=0.95\linewidth]{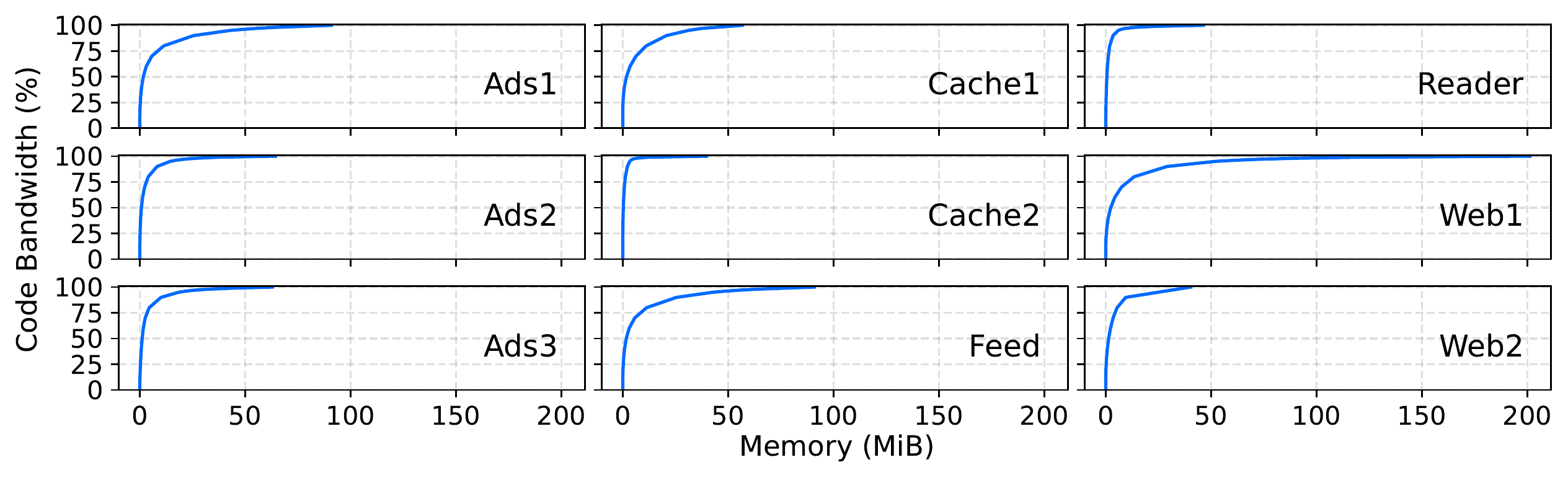}
    \vspace{-0.3cm}
    \caption{Code bw distribution for the nine production workloads using precise L1 \itlb miss events. X-axis represents code pages sorted by their hotness and the Y-axis shows the relative memory bandwidth contribution of the `X' MiB of hottest pages.}
    \label{fig:code-bw-dist}
    \vspace{-0.5cm}
\end{figure*}

As we showed in \refsec{sec:motivation}, CPU frontend pipeline stalls from instruction cache and TLB misses are a significant fraction of the memory system stalls across the nine workloads.
To mitigate the code-fetch overhead, in this section, we look at the code access behavior of cloud workloads at the full system level and propose \uarch{} optimizations. 

To understand the impact of a large code footprint, \reffig{fig:i-bound} shows the percentage of CPU cycles where the CPU core was stalled while waiting for an \itlb miss or an I-cache miss. \web, \insta, and \tao spend over 20\% of their cycles stalling on instruction fetch.
This code fetch overhead is a direct result of the rapid footprint growth of cloud workload binaries due to their high development velocities, as shown in  \reffig{fig:codesize-trend} for \web and \insta workloads. %
Although we use huge pages and code layout optimization techniques~\cite{chen2016autofdo,bolt-anonymous} to reduce \itlb and I-cache misses, we still see fairly high frontend stalls. This is partly because application developers do not control huge page usage and code layout for many dynamically linked libraries; and because the code footprint is too large and still growing. 

To address the code-fetch overhead, our analysis with MemProf reveals that cloud applications execute highly similar code across different CPU cores, in both virtual and physical address spaces, for multi-threaded and multi-process applications alike.

\subsection{Working Set Size for Code}

\begin{figure}
    \centering
    \includegraphics[width=\linewidth]{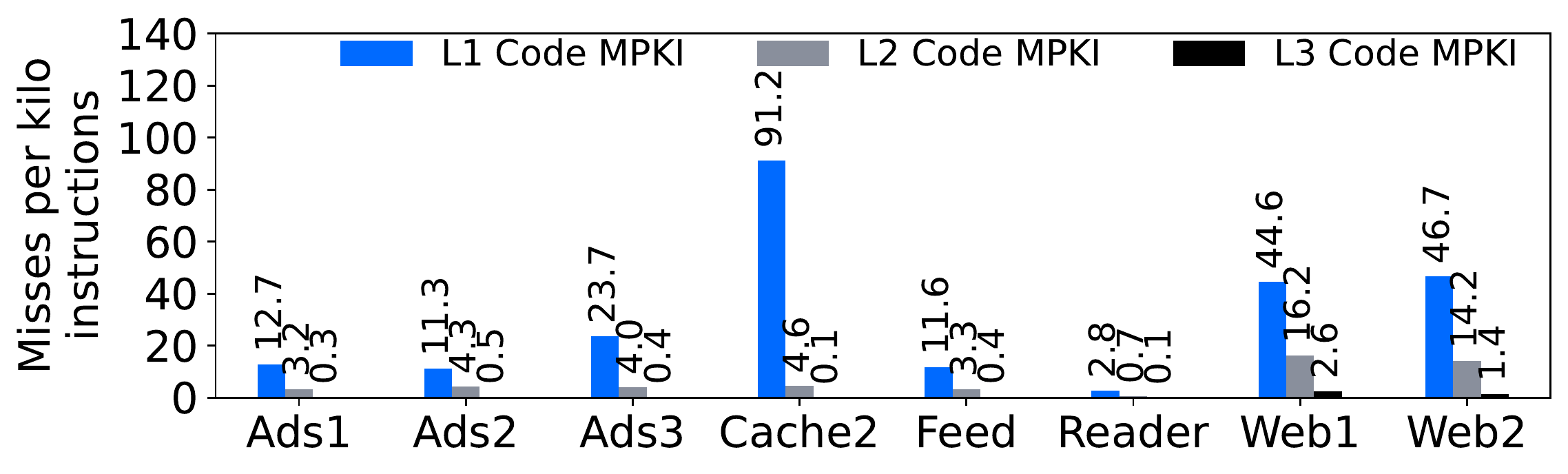}
    \vspace{-0.6cm}
    \caption{L1, L2, and L3 cache code miss rate.}
    \label{fig:code-mpki}
\end{figure}

Next, we will look at the code bandwidth and footprint of the nine workloads to understand why they are significantly code-fetch bound.

Using \memprofcode's L1 \itlb miss event measurements; we generate the code bandwidth of the workloads, as shown in \reffig{fig:code-bw-dist}. We find that all workloads have very large code footprints, with one notable example being the \web workload, which has a code footprint of 125 MiB. 

We further investigate the effects of the large code footprint by measuring the L1-L3 cache code miss rate. \reffig{fig:code-mpki} shows the code MPKI for the three levels of caches. \marginpar{I feel like this is an important observation.}For the nine workloads, we see that the large footprint of cloud workloads results in a high code miss rate across all three levels, with \web and \insta even fetching a significant amount of code lines from  memory.

\hl{Using \reffig{fig:code-bw-dist} and \reffig{fig:code-mpki}, we make three
  important observations: (a) despite the latest processors adopting bigger last-level caches (LLC), certain workloads' huge but sparsely accessed code footprint (see \reffig{fig:code-mpki}) causes diminishing returns in LLC code MPKI improvements. (b) Second, extremely
  large L1 code MPKI values suggest that the active code footprint greatly exceeds
  the L1 code sizes. (c) Finally, workloads like \tao{} can show abnormally high
  L1 cache MPKI and code-fetch bound cycles despite a relatively small code
  footprint. Investigating further, we found that \tao{} suffers from a high
  branch misprediction rate, resulting in increased code accesses and, thus, being significantly more code-fetch
  bound.}

\newcommand{\tx}{$\times$}
It is interesting to note that workloads show a \raisebox{0.5ex}{\texttildelow}10\tx{}  reduction in code MPKI from L2 to L3, which is much higher than the L1-to-L2 reduction. While considering capacity, L2 is often 32\tx{} of L1, while per-core L3 is only about 1.35\tx{} of L2. This suggests that code benefits from shared L3 and effectively uses much more capacity than its per-core quota would suggest. %

\subsection{Code-Sharing Behavior for Cloud Workloads}

\begin{figure*}
  \centering \subfloat[L1 \itlb miss event distribution for the address space of
  \web. Each row is a 2 MiB slice of the address space. Contiguous
  rows represent a contiguous part of the address space.]{ \centering
    \includegraphics[width=0.44\linewidth]{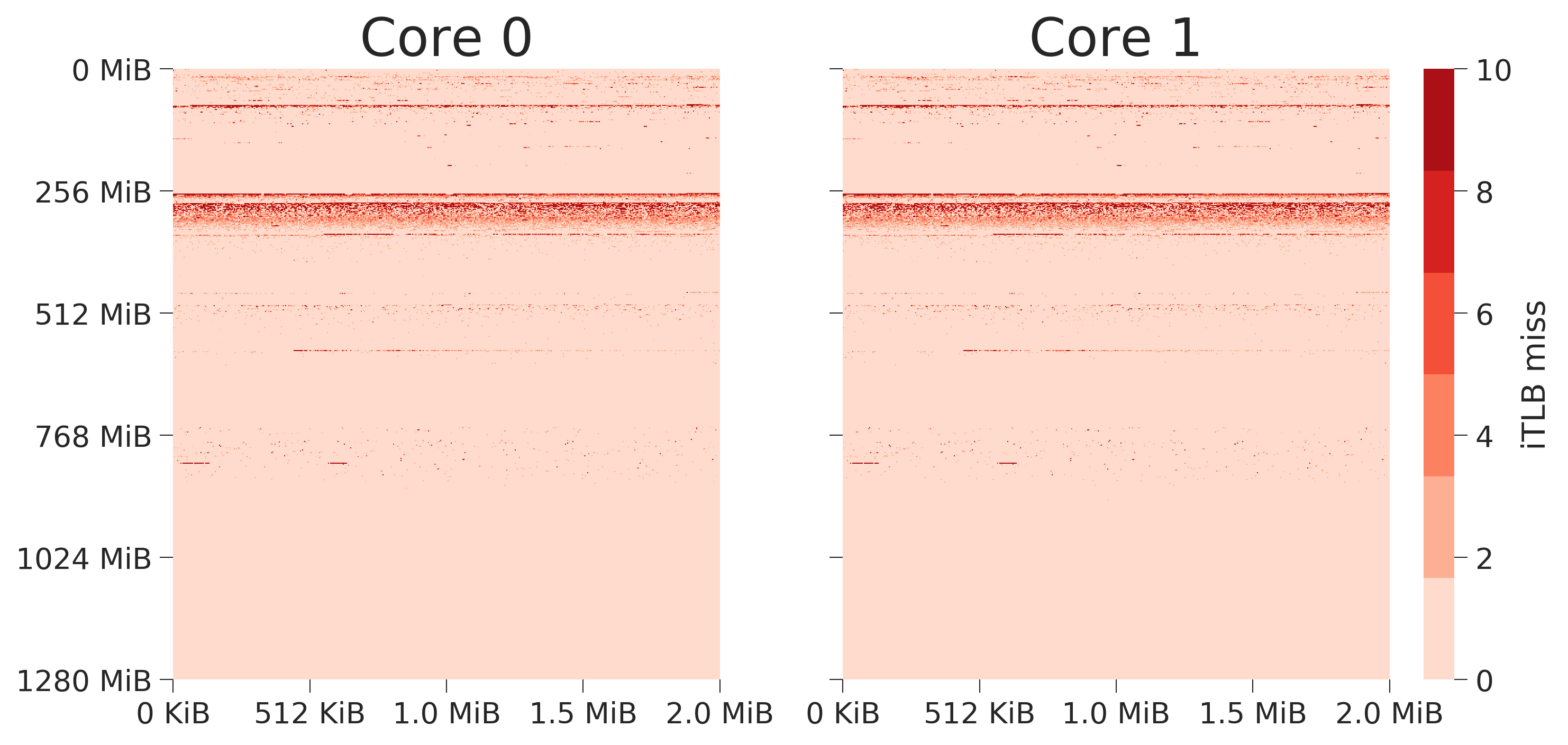}
        \label{fig:itlb-heatmap}
     }
     \hspace{0.8cm}
     \centering
     \subfloat[Hot functions across cores for \web stack sampling.]{
        \centering
        \includegraphics[width=0.42\linewidth]{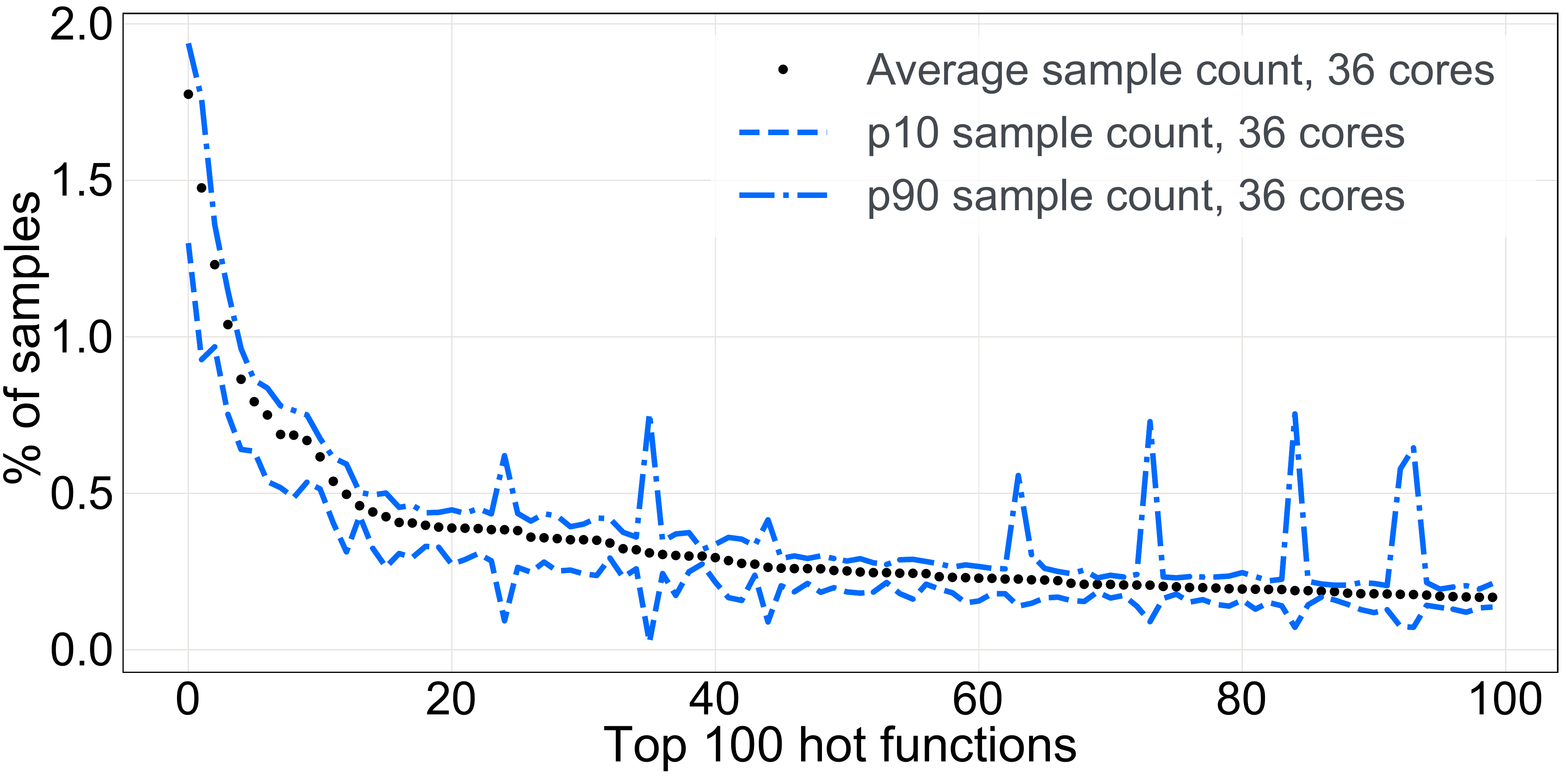}

        \label{fig:top-100-funcs}
     }
    \vspace{-0.1cm}
    \caption{Code sharing across cores using \memprof.}
    \vspace{-0.4cm}
    \label{fig:fleet-ipc}
\end{figure*}

So far, we have looked into the code-fetch overhead of cloud workloads and the resulting performance loss; next, we will present our observation that cloud workloads running across cores share code cachelines, allowing shared cache structures to improve workload's performance. 

To understand if the cores are accessing identical address locations, we used \memprof to sample L1 \itlb miss events for different cores running the same workload. 
\reffig{fig:itlb-heatmap} shows the \itlb miss heatmap for two cores running the \web workload for the first 1.28 GiB of the address space. Across the address space, both cores show very similar code access patterns, suggesting that the cores are accessing common code pages. We test our hypothesis by computing the Pearson correlation coefficient for the two cores; the \web workload shows a correlation value of 0.9997. A Pearson correlation value close to 1 suggests that the L1 \itlb miss events for both cores are highly correlated, and thus, the cores have very similar code access patterns.

{
\renewcommand{\arraystretch}{1.07}

\begin{table}
    \fontsize{8}{10}
    \selectfont
    \centering
    \rowcolors{2}{gray!25}{white}
    \caption{Core 0 \& 1 correlation for L1 \itlb miss events.}
        
    \begin{tabular}{|c|ccc|c|}
        \hline
        \rowcolor[HTML]{d8d8ff}
        {\textbf{Workload}} & {\textbf{Corr. Value}}   &  & {\textbf{Workload}} & {\textbf{Corr. Value}} \\ \hline%
        \web               & 0.9997                      & & \insta         & 0.9881 \\
        \adfinder          & 0.9977                      & & \adranker      & 0.9921 \\
        \adretriever       & 0.9833                      & & \feed          & 0.9977 \\
            &                             & &      &  \\
        \rowcolor{gray!25}
        \doublerown{\memcache b/w\\workload cores}      & \doublerown{0.9947}    & & \doublerown{\memcache b/w\\NIC cores}  & \doublerown{0.9168} \\
        \rowcolor{white}
            &                             & & & \\
        \rowcolor{white}
        \doublerown{\memcache b/w\\workload \& NIC}      & \doublerown{0.0010}    & & \doublerown{\tao}& \doublerown{0.9978} \\
        \dppreader & 0.9887 &&&\\
        \hline
    \end{tabular} 
    \label{tab:correlation-coefficient}
    \vspace{-0cm}
\end{table}
}

\reftab{tab:correlation-coefficient} shows a similar trend for the other workloads. All nine workloads show a high correlation between cores for L1 \itlb miss events. 

For multithreaded workloads, accessing the same pages implies that the CPU cores access common mappings across threads. For the \insta microservice that spawns multiple processes, we use the page table to verify that code page mappings have the same translation across processes.
For \insta{}, we observe common code mappings across processes as all child processes spawn from the same parent.

To confirm that cores execute the same code, we collect stack samples across cores and check if they execute the same functions. 
\reffig{fig:top-100-funcs} shows the average, p90, and p10 weight of the top 100 functions across cores for the \web workload. The L1 I-TLB miss correlation and cores executing identical functions across them confirm that CPU cores execute identical code, leading to our first key observation.
\obs{obs:shared-code-private-data}{Cores run similar code across them.}

Using this observation about cloud workloads, we will propose two new \uarch{} improvements and evaluate their effectiveness \hl{using the \web{} workload. We chose the \web{} workload as it is significantly frontend bound (\reffig{fig:fb-vs-bb}) and runs across a large portion of the fleet.}

\subsection{Reducing I-Cache Misses Using a Shared L2 I-Cache}
\label{sec:shared-l2-cache}
\begin{figure}
    \centering
    \includegraphics[width=0.9\linewidth]{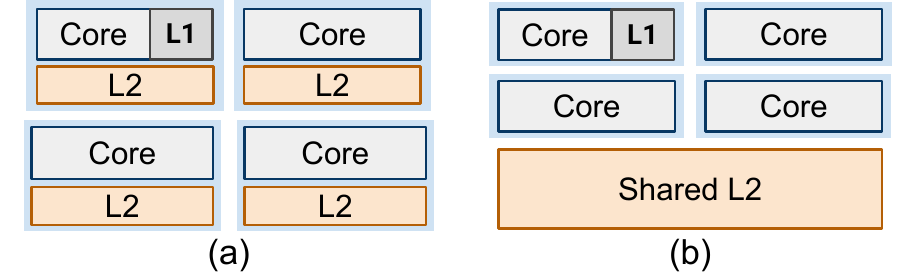}
    \vspace{-0.2cm}
    \caption{Shared cache design with unified L2 cache shared among four cores. (a) Private 1 MiB/core L2. (b) Shared, 1 MiB/core shared L2 (total 4 MiB) cache. }
    \label{fig:shared-l2-arch}
\end{figure}

Since CPU cores share code; with a shared L2 cache, the cores will see an increased apparent cache size for shared cachelines. To take advantage of this, we propose using a shared L2 architecture (\reffig{fig:shared-l2-arch}) where a cluster of four cores shares a unified L2 cache. To keep the hardware cost similar, we propose that the per-core L2 cache size stay similar. 
Thus, a processor resembling the Gen4 platform (\reftab{tab:gen4-config}) should share 4 MiB of L2 cache among four cores (1 MiB/core). In a heterogeneous workload environment, the scheduler can assign multiple threads/processes from a single workload to a core cluster, enabling heterogeneous workloads to take advantage of the  shared L2 cache.

\subsubsection{Evaluation Methodology}
To evaluate the performance improvements from a shared L2 cache, we use Intel's
L2 CAT~\cite{intel-cat} and CDP~\cite{intel-cdp} 
with \web while serving live production traffic.

While simulations can allow us to model the shared cache design closely,
simulations suffer from several limitations. For example, full-system or
system-call emulations have significant performance overhead which changes
the application's runtime behavior. On the other hand, while trace-based simulations have lower
overhead, they cannot accurately capture the timing characteristics,
inter-thread dependencies, or RPCs, making them impractical for cloud workloads.

\subsubsection{Results}

\begin{table} %
    \fontsize{8}{10}
    \selectfont
  \caption{Evaluated system configuration for cache scaling.}
  \label{tab:evalsysconfig}
  \centering
  \rowcolors{2}{gray!25}{white}
  \vspace{-0.2cm}
  \begin{tabular}{|l|l|}
    \rowcolor{gray!25}
    \hline
    \rowcolor[HTML]{d8d8ff}
    \textbf{Parameter} & \textbf{Value}\\
    \hline
    I-L1 Cache                     & 32~KiB/core, private \\\hline
    L2 Cache                     & 2~MiB/core, unified, private                  \\\hline
    LLC                          & 1.875~MiB/core                       \\\hline
  \end{tabular}
  \vspace{0.1cm}
\end{table}

\begin{figure}
    \centering
    \includegraphics[width=\linewidth]{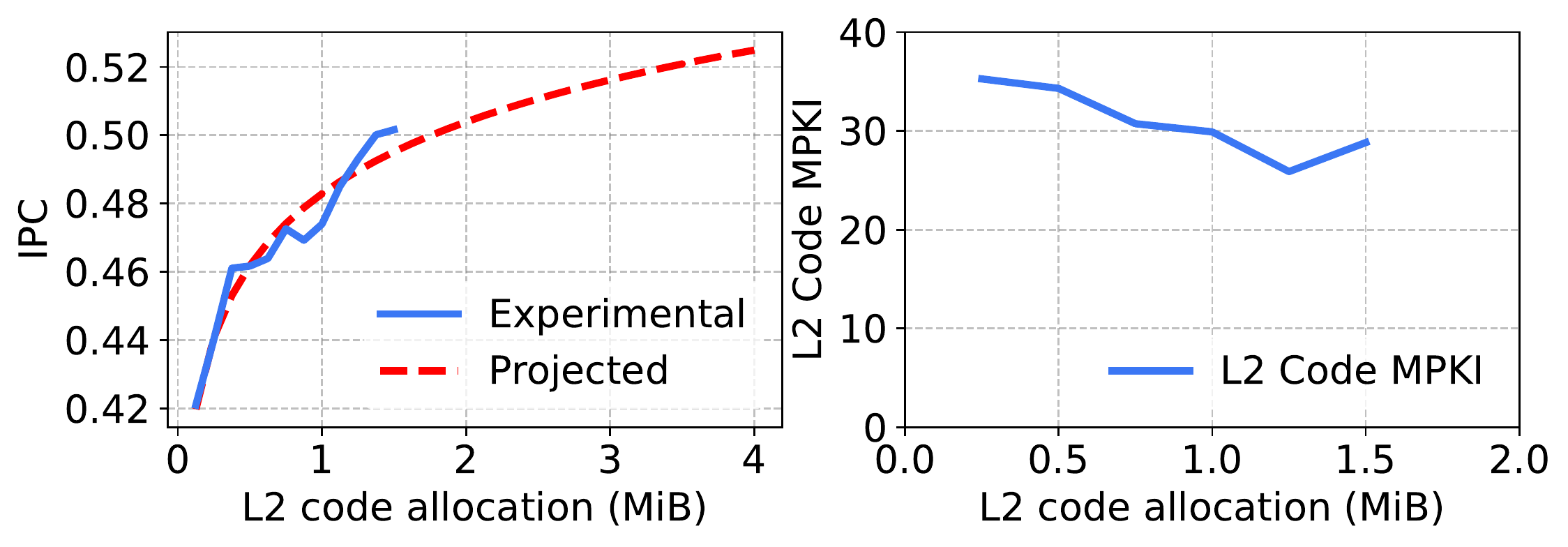}
    \caption{IPC scaling and projection along with L2 code MPKI change for \web with increasing L2 code cache allocation. \web shows an 8.6\% experimental and 9.1\% projected IPC improvement. %
    }
    \label{fig:web-ipc-projection}
\end{figure}

\reffig{fig:web-ipc-projection} shows the measured and projected IPC along with the L2 code MPKI for the \web workload serving live traffic. We use a machine with the cache configuration listed in \reftab{tab:evalsysconfig} and scale the L2 code partition from 128 KiB to 1.5 MiB while keeping the L2 data partition constant at 0.5 MiB (50\% of Gen4 L2\$, based on our study of code-data split in L2 cache). \ignore{To limit the effects of a larger LLC, we limit the LLC size to prevent interference from other cores.} To project performance beyond 1.5 MiB code partition, we extrapolate the experimental data using linear projection in logspace. We chose this projection to model the cache behavior where the performance improves linearly with an exponential increase in cache size~\cite{alameldeen2006ipc}. 

The \web workload shows a real experimental performance improvement of 8.6\% when the cache size increases from 0.5 MiB (50\% of private 1 MiB L2 cache) to 1.5 MiB (50\% of shared 3 MiB L2 cache), corresponding to a 3$\times{}$ increase in apparent I-cache size. For a 4$\times{}$ increase in I-cache size, we rely on projected IPC values, which estimate a 9.1\% increase.

\hl{Thus, based on code sharing across cores and the performance increase from a larger code cache for \web{}, one of the most frontend bound workloads, we observe that cloud workloads benefit from a larger shared L2 cache.}%

\subsection{Reducing \itlb Misses with a Shared L2 \itlb}
\label{sec:shared-itlb}

\begin{figure}
    \centering
    \centerline{\includegraphics[width=\linewidth]{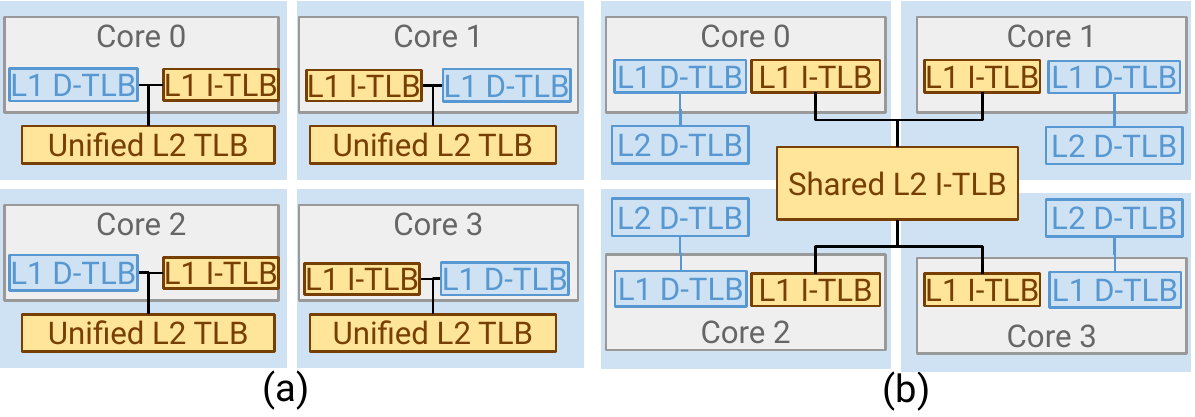}}
    \caption{Shared L2 \itlb design with unified L2 cache shared among 4 cores. (a) Private 1 MiB/core unified L2. (b) Shared, 1 MiB/core Unified L2 (total 4 MiB) cache. }
    \label{fig:shared-itlb-arch}
\end{figure}

\begin{figure}
    \centering
    \centerline{\includegraphics[width=\linewidth]{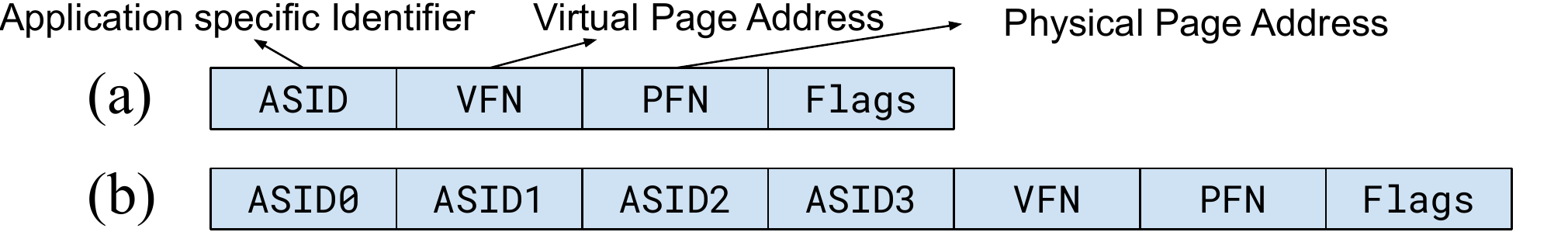}}
    \caption{TLB entry design: (a) private TLB (b) shared TLB.}
    \label{fig:shared-itlb-entry}
\end{figure}

\begin{figure}
    \centering
    \includegraphics[width=\linewidth]{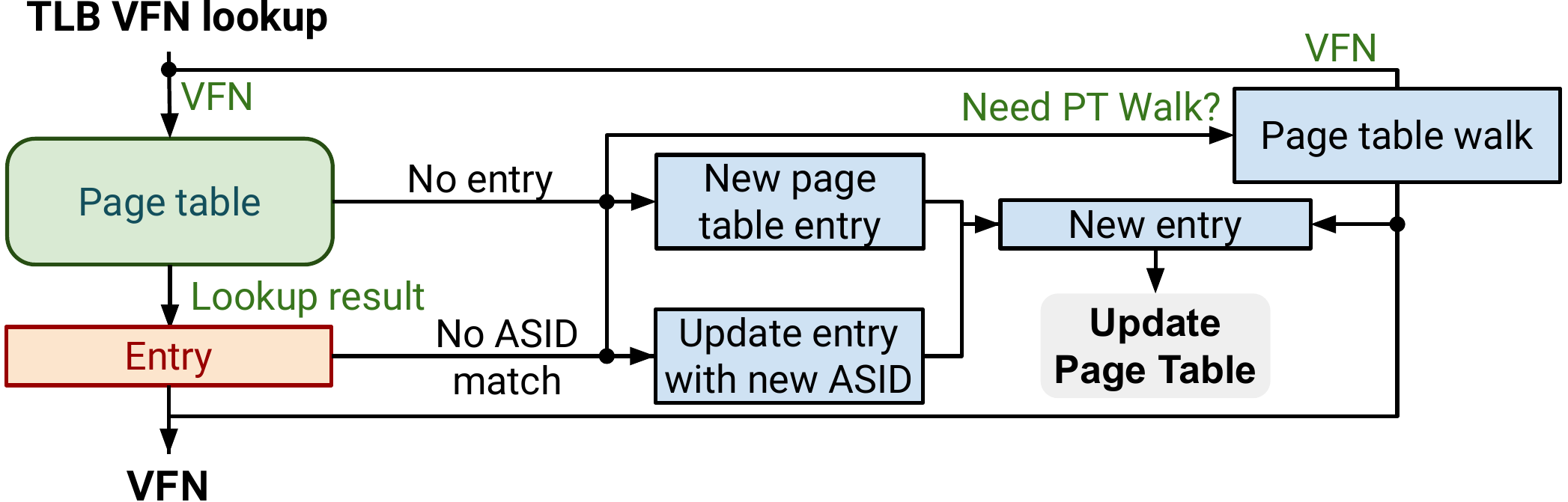}
    \caption{Page table lookup/update for a shared L2 \itlb{}.}
    \label{fig:shared-tlb-flowchart}
\end{figure}

Similar to a shared L2 cache, we propose a shared L2 \itlb to exploit common page mappings for code across cores for cloud workloads. \reffig{fig:shared-itlb-arch} shows a shared L2 \itlb architecture where a cluster of four cores shares the L2 \itlb, but each core still has a private L2 \dtlb. 
\marginpar{Shared libs example?}A shared L2 \itlb allows the cores running identical code to share the TLB entries, making a larger number of TLB entries available to the workload.

Supporting shared \itlb for multithreaded programs requires no additional changes to the TLB entries. Each TLB entry includes an ASID (\reffig{fig:shared-itlb-entry}a) to track which process the entry belongs to. Since each CPU core is part of the same address space in multithreaded programs, the shared unmodified TLB entries can support multiple cores.

For multiple processes across cores, we propose the shared \itlb entries to include multiple ASIDs (\reffig{fig:shared-itlb-entry}b). For our design, we propose 4 processes to share a single L2 \itlb; thus, each TLB entry only needs 4 ASIDs, one for each core.

\reffig{fig:shared-tlb-flowchart} shows the mechanism for a TLB lookup and to install a new entry if it does not already exist. When a translation request comes into the TLB, if the virtual address (VFN) and the entry's application-specific ID (ASID) match, the corresponding physical address (PFN) is returned. If the entry for the VFN does not exist, or the entry's ASIDs do not match the request, the TLB performs a page table walk and adds a new entry or updates the existing page table entry.

\tracing{Finally, as our shared TLB design is similar to shared TLBs in SMT, we expect a comparable security model.}

\subsubsection{Evaluation Methodology}\label{sec:itlb-sharing-validation}

Next, we will validate whether cloud workloads benefit from larger \itlb sizes
by doubling the L1 \itlb entry count.
We cannot do a scaling study similar to \refsec{sec:shared-l2-cache} because the cache partitioning feature is not available for TLBs.
Further, as
previously stated, \uarch{} simulators cannot capture the behavior of a production
environment, making detailed architecture simulation impractical.

\subsubsection{Results}
\begin{figure}
    \centering
    \includegraphics[width=\linewidth]{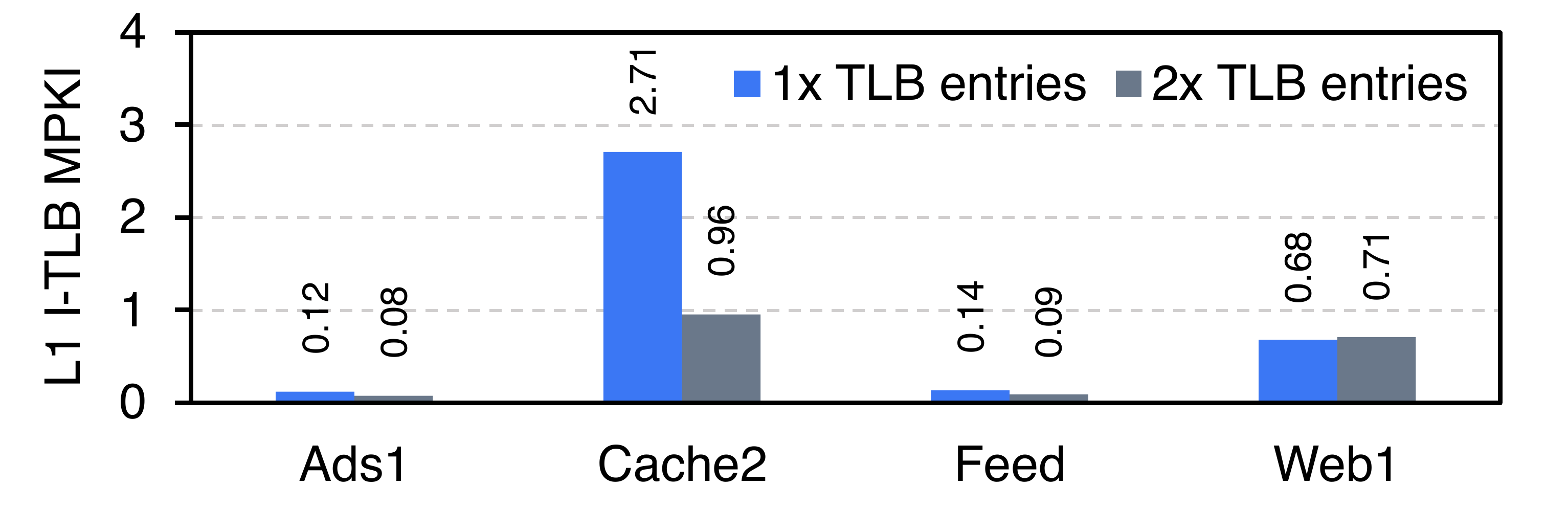}
    \vspace{-0.5cm}
    \caption{I-TLB MPKI with 1$\times$ and 2$\times$ L1 \itlb entries for four cloud workloads.}
    \label{fig:smt-tlb-mpki}
    \vspace{-0.1cm}
\end{figure}

  \reffig{fig:smt-tlb-mpki}
shows the change in L1 \itlb MPKI when the number of entries is doubled, showing
that workloads benefit from an increased \itlb size. To increase the count of
\itlb entries, we measure per-thread L1 \itlb MPKI with simultaneous
multithreading (SMT) on (1$\times$ entries) and off (2$\times$ entries).  

\marginpar{Need to rephrase this.}\hl{Thus, shared code mappings and an increase in performance with more L1 \itlb{} entries show that cloud workloads benefit from a larger, shared L2 \itlb{}. While larger \itlb{} results in a lower MPKI, shared TLB can still suffer from trashing over smaller time intervals if the threads sharing the TLB are in different application phases.} %

\section{Understanding and Improving Memory Bandwidth}\label{sec:mem-bw}
The challenges of memory bandwidth scaling and limited per-DIMM bandwidth have resulted in cloud workloads becoming more memory bandwidth-bound (\refsec{sec:mem-bw-motivation}). 
Since current server processors only support a single memory tier, hyperscalars must populate additional DRAM channels to satisfy the workload's memory bandwidth requirements.
This results in higher power consumption and wasted memory capacity at the datacenter scale.

Recent memory industry trends have resulted in new, \textit{higher} bandwidth memory technologies like Intel MCR~\cite{tomshardware2021skhynix},  HBM, and HB-DIMMs which can be directly connected to CPUs to enable high bandwidths. These technologies offer the potential to explore memory bandwidth tiering as a solution to tackle bandwidth and capacity scaling challenges. It is important to note here that memory-bandwidth contrasts with the memory-latency tiering commonly used today, such as DRAM-Optane tiers. 

In this section, we will study the memory bandwidth characteristics of cloud workloads and look at the potential of high bandwidth memory technologies in improving cloud workloads' performance and server TCO at \company. %

\subsection{Memory Bandwidth Distribution}
\begin{figure*}
    \centering
    \includegraphics[width=0.95\linewidth]{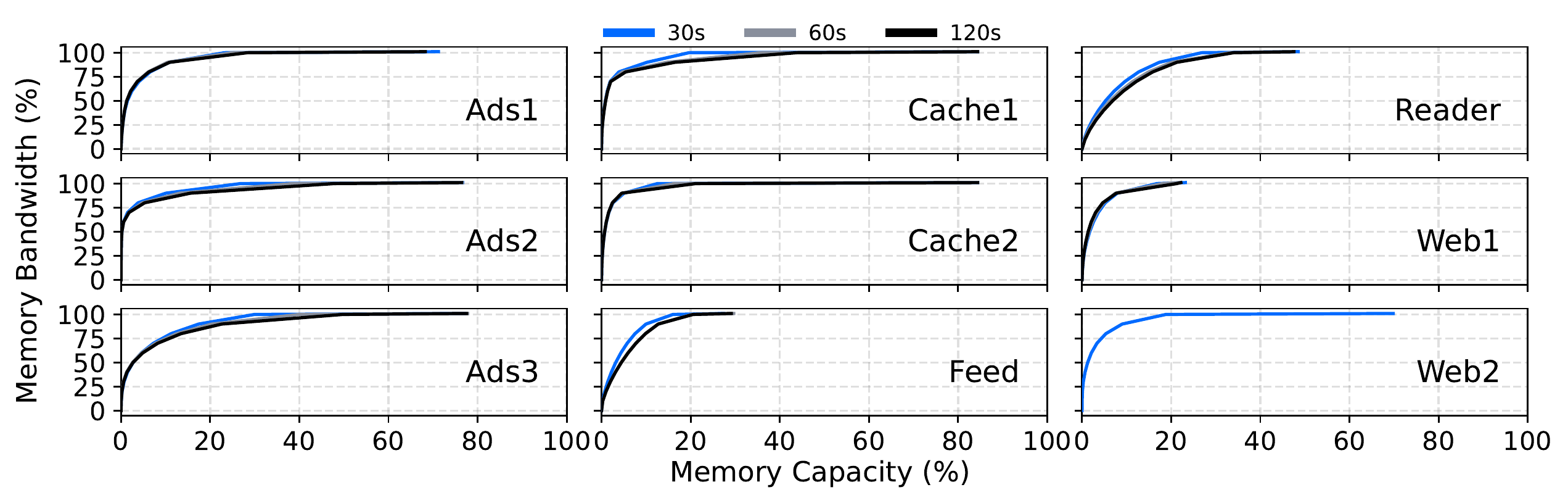}
    \vspace{-0.3cm}
    \caption{Memory bandwidth distribution of the workloads over three different measurement intervals.}%
    \label{fig:bw-dist-bar}
    \vspace{-0.3cm}
\end{figure*}

Using \memprof, we measured the memory bandwidth distribution over 30, 60, and 120 seconds. \reffig{fig:bw-dist-bar} shows the amount of memory contributing to different bandwidth percentiles. Across the nine workloads, we observe that over 30 seconds, the active memory footprint is less than 25\% of the total memory capacity, while the 90\%-tile bandwidth is contributed by less than 10\% of the memory capacity. Moreover, a similar memory bandwidth profile for different measurement intervals suggests that the memory bandwidth distribution remains relatively constant over time, supporting the possibility of memory bandwidth tiering.  %

\memprof, using its LLC demand load miss sampling, enables measuring the relative hit rate of a memory page\ignore{ to compute the memory bandwidth distribution of a workload}. This is unlike current active page tracking techniques, e.g., Kernel Idle Page Tracking (IPT)~\cite{ipt}, which track only active pages with no visibility into how hot an active page is. Since the system does not know the ``hotness'' of a page, these existing techniques cannot measure the memory bandwidth distribution of a workload.

Using \memprof's memory bandwidth distribution experiments, we make our second key observation of cloud workloads' memory behavior:
\obs{obs:mem-bw-dist}{Only a few pages contribute to most of the memory bandwidth over varying time intervals.}
With these insights on memory bandwidth behavior, we will investigate opportunities to improve system performance without substantially increasing cost and power overheads.

\subsection{Memory Bandwidth Tiering}
Memory bandwidth tiering enables servers with memory tiers that efficiently serve memory traffic to the high- and low-bandwidth memory pages. To explore opportunities for memory bandwidth tiering, we will look at memory technologies that support considerably higher memory bandwidth than conventional DDR-DRAM, and study their power requirements and cost overheads. 

\begin{figure}
    \centering
    \includegraphics[width=0.85\linewidth]{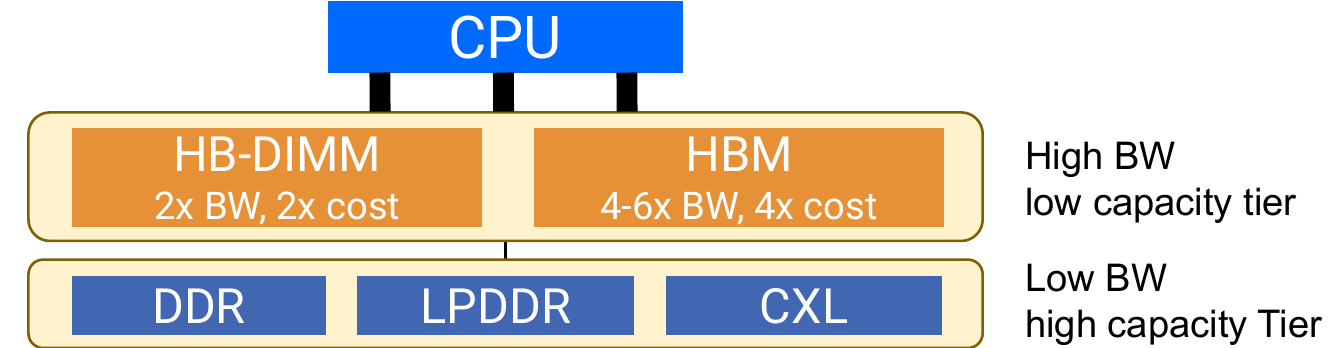}
    \caption{Mem BW tiering architectures for cloud workloads. 
    }
    \label{fig:mem-bw-arch}
\end{figure}

\subsubsection{Opportunities}
Emerging memory technologies like High Bandwidth DIMMs (HB-DIMMs)~\cite{mccall2021high} and High bandwidth memory (HBM)~\cite{standard2013high} enable memory tiers with higher than DDR bandwidth. A higher bandwidth tier in the memory hierarchy enables workloads to meet their bandwidth needs from a small capacity memory while using DDR or CXL-based low-bandwidth, large-capacity tier to meet their memory capacity needs. HB-DIMMs (not to be confused with HBM) provide up to 2$\times$ the bandwidth of conventional DDR5 memory but are expected to consume up to 2$\times$ the power and cost. On-package HBM~\cite{spr-hbm} shows a similar trend (higher bandwidth at the expense of higher power and cost~\cite{li2018performance}). 

\reffig{fig:mem-bw-arch} shows the possible memory bandwidth tiering architecture for cloud workloads based on our observations of the nine representative workloads. While emerging memories support significantly higher memory bandwidth, this comes at increased power consumption and higher cost, making a single high bandwidth tier impractical.

\subsubsection{Evaluation}
To evaluate the possibility of memory bandwidth tiering in datacenters, we will expand memory bandwidth using an HB-DIMM tier, while using a CXL-based memory tier for memory capacity (\reffig{fig:mem-bw-arch}). Please note that while we evaluate a conservative scenario with a higher memory latency far tier (CXL-based), possible memory bandwidth tiers can differ only in memory bandwidth while having similar memory latencies. 
For example, a high-bandwidth tier uses HB-DIMMs, while a low-bandwidth tier uses DDR memories.

\begin{table}
    \fontsize{7}{10}
    \selectfont
    \centering
    \caption{Capacity and theoretical peak memory bandwidth for the three memory bandwidth tiering configurations.}
    \vspace{-0.2cm}
    \rowcolors{2}{gray!25}{white}
    \begin{tabular}{|c|c|c|c|c|c|}
        \hline
        \rowcolor[HTML]{d8d8ff}
        \textbf{Config.} & \textbf{Near Memory}& \textbf{Far Memory}\\\hline
        \textbf{Baseline} & 100\%, 100~GB/s & - \\
        {\textbf{Ideal}} & 100\%, 200~GB/s \textit{(HB-DIMM like)} & {-} \\
        {\textbf{Tiered}} & 37.5\%, 200~GB/s \textit{(HB-DIMM like)} & 62.5\%, 100~GB/s \textit{(CXL-like)}\\
        \hline
    \end{tabular}
    \label{tab:mem-bw-configs}
\end{table}

{
\def\arraystretch{0.8}%
\setlength{\tabcolsep}{0.2em} %

\begin{table}
    \fontsize{7}{10}
    \selectfont
    \centering
    \vspace{0.2cm}
    \caption{Measured relative throughput, mem BW, and relative throughput/cost for the three configurations (\reftab{tab:mem-bw-configs}).}
    \rowcolors{2}{gray!25}{white}
    \vspace{-0.2cm}
    \begin{tabular}{|c|c|c|c|c|c|c|c|}
        \hline
        \rowcolor[HTML]{d8d8ff}
         &  \textbf{} & \multicolumn{2}{c|}{\textbf{Measured}} &   \multicolumn{3}{c|}{} & \textbf{Relative}\\
        \rowcolor[HTML]{d8d8ff}
         &  \textbf{Relative} & \multicolumn{2}{c|}{\textbf{BW (GiB/s)}} & \multicolumn{3}{c|}{\doublerown{\textbf{Cost (relative)}}} & \textbf{Tput.} \\\cline{3-7}
        \rowcolor[HTML]{d8d8ff}
        \triplerown{\textbf{Config}} &  \textbf{Tput.} & \textbf{\hspace{0.3cm}Near\hspace{0.3cm}} & \textbf{Far} & \textbf{\hspace{0.1cm}Near\hspace{0.1cm}} & \textbf{Far} & \textbf{Total} & {\textbf{/cost}} \\\hline
        \textbf{Baseline}  & 1 & 67.81 & - & 1.0 & - & 1.0 & 1\\
        \textbf{Ideal}  & 1.55 & 108.41 & - & 2.0 & - & 2.0 & 0.73 \\
         &  &  &  & 0.75 & 0.625  &  & \\
         \rowcolor{gray!25}
        \doublerown{\textbf{Tiered}}& \doublerown{1.47} & \doublerown{84.60} & \doublerown{19.22} & \textit{(2$\times$37.5\%)} & \textit{(1$\times$62.5\%)} & \doublerown{1.375} & \doublerown{1.13}\\
        \hline
    \end{tabular}
    \label{tab:mem-bw-results}
\end{table}
}

We use the three configurations listed in \reftab{tab:mem-bw-configs} for our evaluation. The baseline configuration resembles current servers deployed in the datacenters at a hyperscalar. In contrast, the Ideal configuration shows the maximum possible performance improvement using a high-bandwidth-only tier. Finally, the Tiered configuration allocates about 1/3 of the capacity using the high-bandwidth tier that resembles HB-DIMMs, while rest of the capacity is allocated on far memory that resembles CXL memory. The tiered configuration uses Maruf et al.'s~\cite{maruf2022tpp} page placement technique.

We use a dual-socket server with DDR-5 DIMMs and configurable channel counts to allocate the memory bandwidth. Only one socket has its CPU cores enabled, with the memory connected to this socket referred to as \textit{Near Memory}. Memory connected to the socket with all its CPU cores disabled is referred to as \textit{Far Memory}. HB-DIMMs are expected to have DDR-5 like latencies, so we model HB-DIMMs using DDR-5 based near memory. Likewise, CXL memories are expected to have latency and bandwidth characteristics similar to NUMA links between the sockets~\cite{maruf2022tpp} (Intel UPI). Thus, \textit{Far Memory} represents a CXL memory tier. 

\subsubsection{Results}
\begin{figure}
    \centering
    \includegraphics[width=\linewidth]{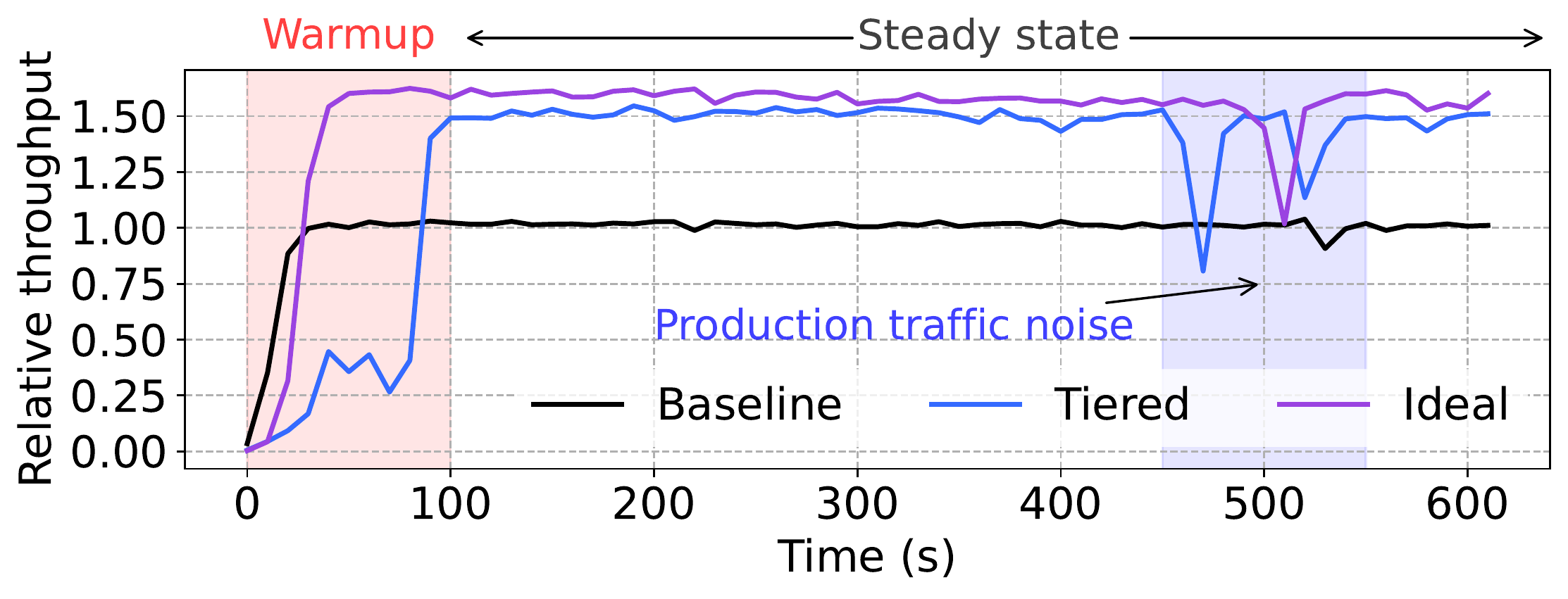}
    \vspace{-0.6cm}
    \caption{Relative throughput of the \dppreader workload for the three memory configurations (\reftab{tab:mem-bw-configs}).}
    \label{fig:req-rate-time-series}
\end{figure}

\reffig{fig:req-rate-time-series} shows the relative throughput of the \dppreader{} microservice serving live production traffic for the three configurations. At the steady state ($t > 100~\mathrm{s}$), the Tiered memory configuration achieves 1.46$\times$ better throughput than the Baseline and within 6.32\% of the Ideal configuration. Out of the three configurations, the Tiered configuration takes the longest to warm up from the page migration overhead between near and far memories, which the Baseline and Ideal configurations do not have. \hl{Due to the limited and geo-restricted availability of the evaluation server machine, we only evaluate the \dppreader{} microservice, which is the most backend-bound workload out of the nine representative workloads (\reffig{fig:fb-vs-bb}).}

\reftab{tab:mem-bw-results} lists the three configurations' measured memory
bandwidth and relative throughput/cost
As HB-DIMMs are not commercially available, for our TCO analysis, we use an early assessment of HB-DIMM's cost as 2$\times$ that of DDR-5 DIMMs. The tiered memory configuration provides a 13\% improvement in performance/cost over the Baseline, and a 54.8\% improvement over the Ideal configuration, making it the most cost-efficient memory hierarchy, confirming our memory bandwidth observations using \memprof.

\section{Understanding Memory Latency}\label{sec:mem-lat}

\begin{figure}
    \centering
    \includegraphics[width=\linewidth]{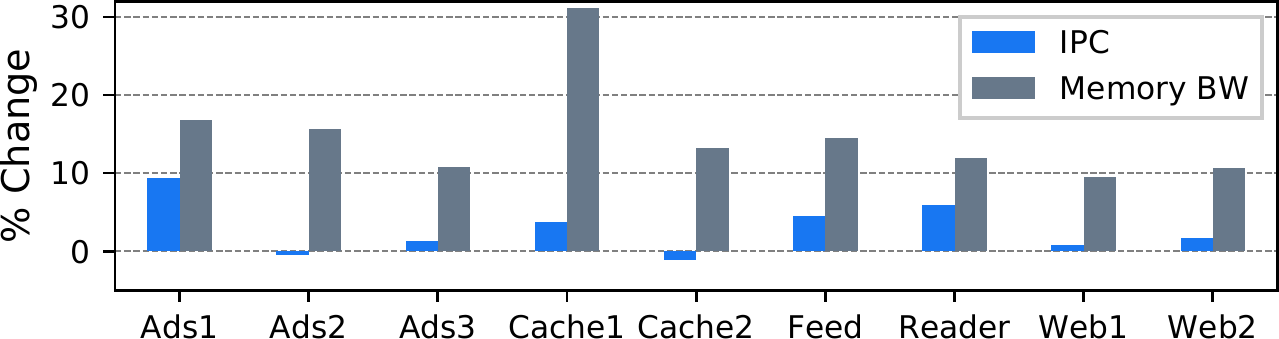}
    \vspace{-0.5cm}
    \caption{IPC and mem BW change with L2 prefetchers enabled.} %
    \label{fig:ipc-mem-bw-l2-pf}
\end{figure}

Memory latency is another significant concern for cloud workloads~\cite{ayers2018memory}, with server processors often relying on hardware prefetchers to alleviate the performance penalty. \AB{Should we also mention that SW prefetching is not a standard practice in our environment for most workloads and generally across cloud workloads because of difficulty in implementing them effectively with large code footprints.} \ignore{\reffig{fig:emon-trend} showed that a large percentage of cycles had at least one outstanding instruction waiting for the memory to respond.} With cloud microservices operating under strict request latency SLO constraints, workloads must limit CPU utilization to meet the SLO guarantees, resulting in wasted resources, \hl{as discussed in \refsec{sec:mem-bw-motivation}}.

A common solution to hide memory latency, the hardware prefetchers, work by prefetching soon-to-be-used cachelines, thus reducing the overall memory latency. While prefetchers hide some memory latency, they also significantly increase  memory bandwidth consumption. \reffig{fig:ipc-mem-bw-l2-pf} shows the increase in IPC and the corresponding increase in memory bandwidth with L2 prefetchers turned on. Several workloads show a significant increase in memory bandwidth consumption (e.g., for \memcache, it increases by 31\%), signaling prefetcher inefficiencies. \hl{While we present data for L2 prefetchers, from our experiments, we observe that L1 prefetchers have much higher efficiencies; thus, we omit them from our study.}

\hl{Although modern hardware prefetchers monitor bandwidth utilization and throttle to reduce inefficiencies~\cite{heirman2018near}, we observe that workloads like \adfinder{} and \dppreader{} still suffer from a large memory bandwidth overhead.}

Using \reffig{fig:ipc-mem-bw-l2-pf}, we make several important observations: (a) while some workloads show negligible IPC improvement, we keep L2 prefetchers enabled as the servers have enough memory bandwidth headroom for the listed workloads. In the past, with limited memory bandwidth, turning L2 prefetchers off has resulted in performance improvements. (b) Workloads with predictable memory access patterns (e.g., CPU-based inference) like \adfinder{} show a significant IPC improvement, suggesting efficient hardware prefetchers are important in improving  performance.

Next, we will look into reasons for prefetcher inefficiencies and future opportunities to mitigate them using production memory access traces.
\subsection{Hardware Prefetcher's Accuracy and Coverage}
\begin{figure}
    \centering
    \includegraphics[width=1\linewidth]{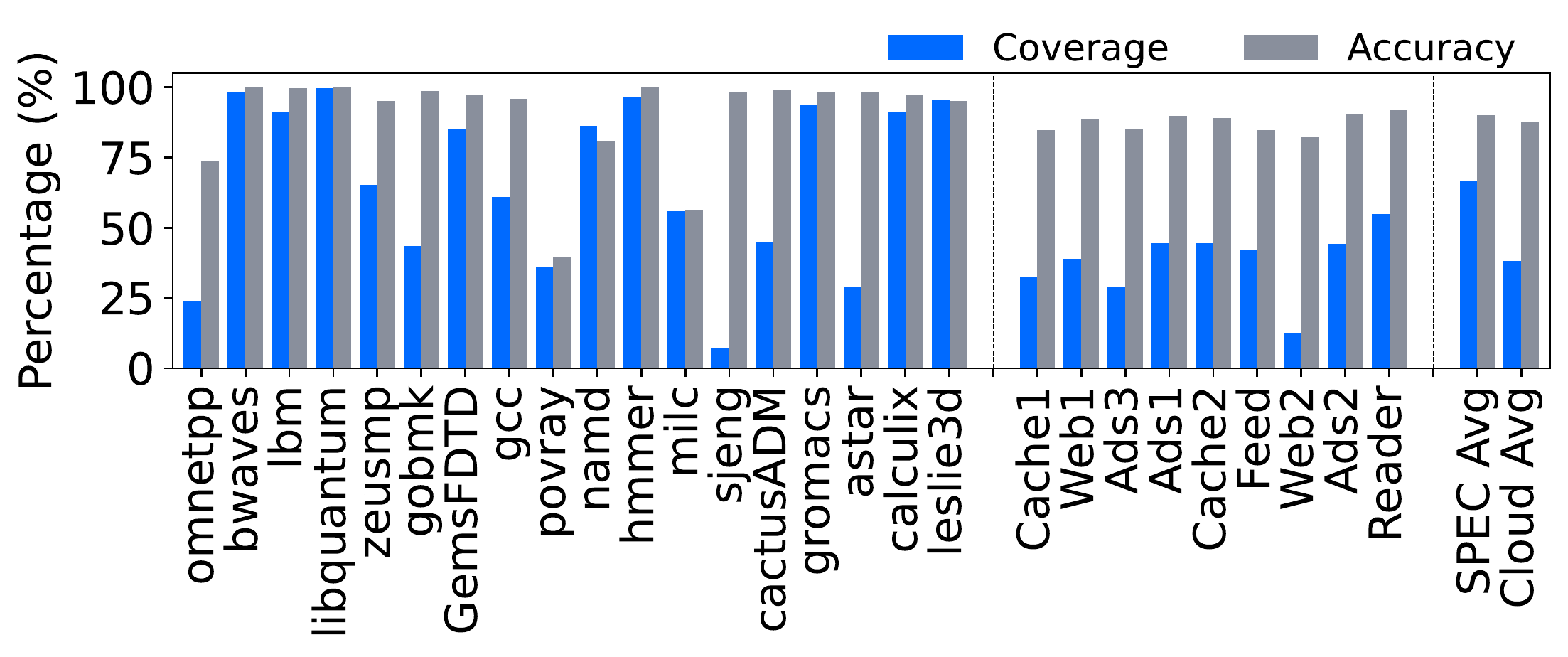}
    \vspace{-0.5cm}
    \caption{L2 hardware prefetcher accuracy and coverage.}
    \label{fig:hw-pf-accuracy-coverage}
\end{figure}

\reffig{fig:hw-pf-accuracy-coverage} shows the accuracy and coverage of L2 hardware prefetchers for the nine workloads and SPEC 2006 CPU benchmark. Most cloud workloads show very low coverage (<50\%) but are relatively accurate (>75\%). This tells us that the L2 hardware prefetcher cannot predict a large fraction of access patterns of the workloads. Further, it is interesting to note that cloud workloads on average,  have considerably lower prefetcher coverage compared to the SPEC benchmark, 38.1\% vs. 67.0\%.

While there have been proposals to improve hardware prefetchers~\cite{litz2022crisp,ayers2020classifying}, research on hardware prefetchers for cloud workloads is limited by access to these workloads \tracing{and the difficulty of running them in a simulator.} 
\tracing{Studying their memory access traces is an alternative, but today's memory tracing tools have a significant runtime overhead, making them impractical for production workloads.}

\subsection{Efficient Memory Tracing for Future Architecture Research}
\tracing{Tracing memory accesses in production workloads poses several challenges, necessitating the development of a new tracing technique. In response, we present \memprof.Trace, a low-overhead memory access tracing tool. 
Next, we look at the challenges in tracing memory accesses in a production environment followed by the description and evaluation of \memprof.Trace.}

\subsubsection{Challenges of Tracing Memory}
\tracing{The two major challenges are:}

    \textit{Performance overhead.} Common memory tracing tools use dynamic instrumentation to trace accesses but add significant execution overhead to the application.
    \ignore{This overhead often slows down the application significantly (e.g., 5-10$\times$ for Intel's Memory Latency Checker~\cite{intel-mlc} in our experiments).}  The effect of workload slowdown is different across the different production workloads we tried to trace, with common causes being: (a) the kernel scheduler schedules some other thread on the core, (b) requests timeout, resulting in retries or failures, and (c) the load balancer throttles machine load.
    
    \textit{Dynamic application phases.} Tracing memory access at smaller granularities to limit overhead poses a challenge with workloads that exhibit diverse and short-lived application phases. Examples of these phases are: parsing a request, communicating with other microservices, or querying a database. Thus, collecting a single small trace (millions of accesses) would not be enough to get representative behavior of the workload.

\subsubsection{{MemProf.Trace} -- Low-overhead Memory\\Tracing}
To overcome these challenges, we built a PIN Tool that periodically attaches itself to a process, records memory access for a configurable duration (in the order of microseconds), and detaches itself to let the workload continue. We collect several such traces from multiple hosts and stitch them together to create a representative workload trace. \hl{Thus allowing low-overhead tracing where services do not violate their request-level latency guarantees.}

\begin{table}
    \fontsize{8}{10}
    \selectfont
    \centering
    \rowcolors{2}{gray!25}{white}
    \caption{Measured (Prod.) and simulated (Trace) results.}
    \begin{tabular}{|c|c|c|c|c|c|c|}
        \rowcolor[HTML]{d8d8ff}
        \hline
         & \multicolumn{3}{c|}{\textbf{L1D Hit Ratio}}& \multicolumn{3}{c|}{\textbf{R:W Ratio}}\\\cline{2-7}
        \rowcolor[HTML]{d8d8ff}
        \doublerown{\textbf{}} & \textbf{Prod.} & \textbf{Trace}  & \textbf{Error}  & {\textbf{Prod.}} & \textbf{Trace} & \textbf{Error}\\\hline
        \memcache & 0.93 & 0.88  & 5.38\% & 1.84 & 1.92 & -4.34\%\\\hline
        \feed & 0.95 & 0.93  & 2.11\% & 2.14 & 2.20 & -2.8\%\\\hline
        \web & 0.94 & 0.90  & 4.25\% & 1.72 & 1.67 & 2.3\%\\\hline
        
    \end{tabular}
    \label{tab:hit-rate}
\end{table}

\begin{figure}
    \centering
    \includesvg[width=\linewidth]{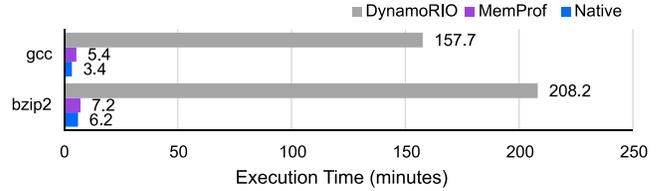}
    \vspace{-0.5cm}
    \caption{Tracing overhead, \memprof vs DynamoRIO.}
    \label{fig:tracing-overhead}
    \vspace{-0.05cm}
\end{figure}

\subsubsection{Overhead and Trace Validation}
\tracing{\textbf{Performance Overhead.} We compare \memprof.Trace against state-of-the-art memory tracer, DynamoRIO~\cite{bruening2012transparent}, and native execution for gcc and bzip2 from the SPEC 2006 CPU benchmark. \reffig{fig:tracing-overhead} shows that compared to \memprof, DynamoRIO has orders of magnitude higher performance overhead. For the experiment, we configured \memprof's sampling rate at 100~ms every 5~s. Due to long execution times, we limit performance evaluation to two workloads.}

\textbf{Trace Validation.} To verify the accuracy of the traces, we used a simple cache simulator with the same cache architecture as the production servers and verified the L1D hit rate. 
\reftab{tab:hit-rate} shows the measured and simulated hit rate for the L1-D cache and the R:W ratios. All workloads show a small error, signifying that the traces are accurate for studying the memory behavior of the workloads. 

\tracing{We will make these traces available to the community in the future to help advance hardware prefetcher research.}

\section{Related Works}
\label{sec:related}

Some previous works have analyzed data center workloads code and data behavior
to understand their characteristics. \ignore{These works have looked at the
  breakdown of stalls and made observations related to the growth of the code
  footprint.} To the best of our knowledge, none of the prior works have looked
at code sharing across threads and processes, memory bandwidth distribution of
production workloads, and the impact of hardware prefetchers on memory bandwidth
and performance.

Ayers et al.~\cite{ayers2018memory} analyzed Google's web search workload, observing large code footprints and high MPKI for caches. They proposed a large, low-latency L4 cache. In contrast, we suggest sharing L2 I-Cache and L2 I-TLB among four-core clusters, improving performance with minimal area and power increase.
Kundu et al.~\cite{kundu2003case} also studied
a shared L1 cache design, but using an OLTP benchmark. \chpar{Similarly, Alves
  et al.~\cite{alves2009investigation} studied shared L2 cache (data+code) using
  the NAS benchmark. HPC benchmarks have significantly different performance
  characteristics, so they observe significant performance regression from cache
  sharing. Tam et al.~\cite{tam2007managing} use software-based L2
  cache partitioning to avoid contention between cores which resembles Intel's CAT~\cite{intel-cat}}.

Kanev et al.~\cite{kanev2015profiling} investigated ``datacenter tax'' overhead at Google, identifying I-cache as a significant bottleneck due to growing code footprints in cloud workloads. They observed low memory bandwidth consumption resulting from overprovisioned machines, contrasting our findings that memory bandwidth is increasingly concerning. Nevertheless, \memprof insights can help future hardware prefetcher research, offering possibilities for overall performance improvements.

SoftSKU~\cite{sriraman2019softsku} examines Meta's cloud workloads, understanding OS and hardware behavior using "Soft SKUs" by adjusting core/uncore frequencies, LLC prioritization, and active core counts. They observe similar IPC, cache, and TLB MPKIs. Accelerometer~\cite{sriraman2020accelerometer} studies Meta's cloud workloads, finding most cycles spent on non-core application tasks, such as compression and serialization.

Some earlier
works~\cite{trancoso1997memory,barroso1998memory} have also looked at \uarch{}
improvements for cloud workloads. A few workloads have analyzed the commercial
benchmarks using similar techniques, e.g., PIN Tool~\cite{jaleel2007memory} and
top-down analysis~\cite{yasin2014top}.

\hl{For memory traces, we ensure that they accurately represent production
  behavior which we validate using \uarch{} statistics. Ranganathan et
  al.}~\cite{google-traces} \hl{ collected traces from production using dynamoRIO~\cite{bruening2012transparent}. Our
  approach is complementary to theirs.} 
\tracing{Payer et al.~\cite{payer2013lightweight} leverage 64-bit register sizes while executing 32-bit code to trace memory accesses, implementing a lightweight memory tracing solution. However, 32-bit code has become increasingly uncommon in modern computing. Daptrace~\cite{lee2020lightweight} employs page table access bit sampling to detect hot objects but does not record individual memory accesses. 
HMTT~\cite{bao2008hmtt} uses a dedicated hardware board for DIMM monitoring without application changes. However, HMTT requires specialized hardware and server modifications, often impractical in data centers. In contrast, MemProf avoids hardware changes, ensures minimal intrusiveness, and achieves low performance overhead.}

Some recent works have looked into the memory prefetcher behavior of SPEC and
cloud workloads. Litz et al.~\cite{litz2022crisp} present critical slice
prefetching (CRISP) to prefetch hard-to-predict loads\ignore{ using a new
  instruction prefix that increases the instruction's priority in the
  instruction scheduler}. \chpar{Jamilan et al.~\cite{jamilan2022apt} and
  I-SPY~\cite{khan2020spy} rely on Intel's LBR and dynamic execution information
  to optimize data and instruction prefetching, respectively.}  \ignore{CRISP
  uses a hardware-software technique to identify and tag instructions.} Ayers et
al.~\cite{ayers2020classifying} study SPEC and Google workloads and present an
automated way of classifying memory access patterns for software-based
prefetching. \chpar{Hashemi et al.~\cite{hashemi2018learning} use machine
  learning to study memory traces. \memprof enables low-overhead, accurate traces
  complementing their techniques. These works are orthogonal to ours and can be
  used together to optimize different aspects of cloud workloads.}

\section{Conclusion}
\label{sec:conclusion}

Memory subsystem bottlenecks are one of the primary reasons for stagnating IPC on cloud workloads. Datacenters are a significant portion of the server processor market, but benchmarks used for CPU designs often do not represent their characteristics. In this work, we present a detailed characterization of \company's cloud workloads using \memprof and use the findings that cores share code cachelines to propose shared L2 cache and shared L2 I-TLB caches. Next, we look into the memory bandwidth behavior of live cloud workloads and evaluate memory bandwidth tiering to achieve throughput and TCO improvements. Finally, we look into ways to improve memory latency. We collect production memory traces and verify their accuracy for future memory prefetcher research. %

\bibliographystyle{plain}
\bibliography{references}

\end{document}